\documentstyle[seceq,epsf,wrapft]{ptptex}

\title{Quantum Tunneling in Half-Integer Spin Systems}

\author{Seiji {\sc Miyashita}$^{1)}$\footnote{E-mail address: 
miya@spin.t.u-tokyo.ac.jp}
 and 
Naoto {\sc Nagaosa}$^{1,2)}$\footnote{E-mail address: nagaosa@appi.t.u-tokyo.ac.jp}
}

\inst{$^1$ Department of applied Physics, University of Tokyo,
Hongo, Tokyo 113-8656\\
$^2$ Correlated Electron Research Center (CERC), 1-1-4 Higashi, Tsukuba 
305-0046, Japan }

\recdate{
\today
}

\abst{
Motivated by the experimental observations of resonant tunnelings
in the systems with half-integer spin, such as V$_{15}$ and Mn$_4$, we study
the mechanism of adiabatic change of the magnetization in
systems with the  time-reversal symmetry. 
Within the time-reversal symmetric models,
effects of several types of perturbations are investigated.
Although tunneling between the ground states is suppressed in a simple Kramers doublet,
we show that the nonadiabatic
transition governed by the Landau-Zener-St\"uckelberg mechanism
occurs in many cases due to the additional degeneracy of the ground state.
We also found more general cases where LZS mechanism can not be 
applied directly even the system shows a kind of adiabatic change of the magnetization.
}

\def\beq{\begin{equation}}
\def\eeq{\end{equation}}
\def\beqa{\begin{eqnarray}}
\def\eeqa{\end{eqnarray}}
\def\beqp{\begin{equation}\left\{\begin{array}{ll}}
\def\eeqp{\end{array}\right.\end{equation}}
\def\bfg{\begin{figure}}
\def\efg{\end{figure}}
\newcommand{\mbold}[1]{\mbox{\boldmath $ #1 $}}

\begin{document}

\maketitle

\section{Introduction}

Recently the dynamics of nanoscale molecular magnets has 
attracted much interest. 
In particular, the magnetization processes in some of molecules 
show step-wise structures of hysteresis in the sweeping magnetic field. 
This phenomenon is attributed to the quantum mechanical resonance 
at level crossing points in discrete energy structure of the small magnets,
and is called {\it resonant tunneling}.
The low-energy magnetic properties of such molecules is usually given by a model of 
a single large spin. 
For example, the system of Mn$_{12}$\cite{mn12} or Fe$_8$\cite{fe8,feW} is represented by
\beq
{\cal H}_0=-D(S^z)^2-HS^z, \quad S=10, \quad D>0.
\label{eq1}
\eeq
Level-mixing interactions such as
\beq
{\cal H}_1=C((S_i^+)^4+(S_i^-)^4)
\quad {\rm or} \quad
{\cal H}_1=E((S_i^x)^2-(S_i^y)^2).
\eeq
cause hybridization of the degenerate states of different values of
 the magnetization, and
the resonance between the two states takes place, i.e., the resonance tunneling.
The hybridization causes an energy gap, i.e., the tunneling gap.
In particular, in the case of integer spin $S$,
the hybridization occurs between two-fold degenerate ground states 
with the total magnetization $M=\pm S$ at $H=0$. 
This tunneling phenomena has been studied by 
a kind of path-integral method\cite{CG} which has provided 
various important informations.
There the tunneling probability is expressed by a 
form of path integral representation, e.g., 
\beq
P= \langle\pi|e^{-\beta{\cal H}}|0\rangle,
\label{P}
\eeq
where $H=K_x(S^x)^2+K_yS^y,\quad K_x>K_y>0$ and $|\pi\rangle$ and $|0\rangle$
denote the states directing $\pm z$ directions (the easy axis), respectively.

However, in this picture of a single large spin model, 
the tunneling in half-integer spin particle is known to be suppressed 
due to the interference of the Berry phase\cite{HS,LDG,Garg}.
This fact is originated from the time-reversal symmetry and it corresponds to 
the Kramers theorem that all the energy levels must
be at least doubly degenerate in half-integer spin systems.
In the context of eq.(\ref{P}), the contributions from the pair of the two
paths related by the time-reversal operation cancel with each other
and
\beq
\langle\pi|e^{-\beta{\cal H}}|0\rangle=0,
\label{PI}
\eeq
when $S=$ half-integer, from which one may conclude
the absence of the tunneling gap.

Generally, systems have the time-reversal symmetry when the hamiltonian of the system
consists of products of even number of spins, e.g.,
the usual two spin interactions, etc.
Therefore, in systems of odd number of spins, the eigenstates must be at least 
doubly degenerate at $H=0$, i.e. the Kramers theorem.
However, in many half-integer spin systems, e.g. V$_{15}(S=3/2)$\cite{V15,V15-2}
and Mn$_4(S=9/2)$\cite{Mn4}, temperature independent
relaxation phenomena have been found, which suggests the existence of 
a tunneling splitting in half-integer systems.
 
In this paper, we consider this tunneling problem in time-reversal symmetric
systems. The key point is the fact that    
the realistic molecule consists of many atoms and many degrees of freedom
exist. There the magnetic property can not be simply expressed by a single large spin.
We find that the system can show tunneling behavior making use of 
additional degrees of freedom, even if each level are the Kramers doublet.

The existence of the tunneling gap causes an avoided level crossing structure
of the energy level as a function of the magnetic field $H$
\beq
{\cal H}_{\rm Z}=-H\sum_iS_i^z,
\label{Zeeman}
\eeq
and we expect the adiabatic motion of the
magnetization when we sweep the magnetic field.
As time-dependent phenomena, the magnetization adiabatically follows 
the ground state value if the field-sweeping is slow. 
There the sign of  the magnetization changes near $H=0$, which is the adiabatic motion.
If we sweep $H$ fast, then generally the nonadiabatic transition 
occurs\cite{Landau,Zener,St,nonad1,KN98,Ao,miya1,miya2,th6,th7,GC},
and the transition probability of the  nonadiabatic transition was
given by Landau, Zener and St\"uckelberg\cite{Landau,Zener,St} as
\beq
p=\exp(-{\pi(\Delta E)^2\over 2v}),
\label{LZS}
\eeq
where $v$ is the sweeping rate of the field $dH/dt$, and 
$\Delta E$ is the energy gap at the avoided level crossing 
structure.
This transition rate plays an important role in the resonant tunneling
phenomena in nanoscale molecular magnets such as Mn$_{12}$ and 
Fe$_8$\cite{feW,miya1,miya2,th6,th7,GC}.
In the present paper we point out that the various types of 
avoided level crossing structures are possible in half-integer spin systems
which are time-reversal symmetric at $H=0$, and 
also we study properties of adiabatic changes of the magnetization.

\section{Effects of magnetization-nonconserving matrix elements}

In this paper, we mainly consider the Heisenberg model with odd number of spins
as the unperturbed system.
In order to change the value of the magnetization, the
system must contain some terms which violates the conservation of the
magnetization.
That is, the hamiltonian must not commute with the total magnetization.
We will see that some of such systems have the avoided level crossing structures.

\begin{figure}
\begin{eqnarray}
\begin{array}{ll}
\epsfxsize=6.0cm \epsfysize=6.0cm \epsfbox{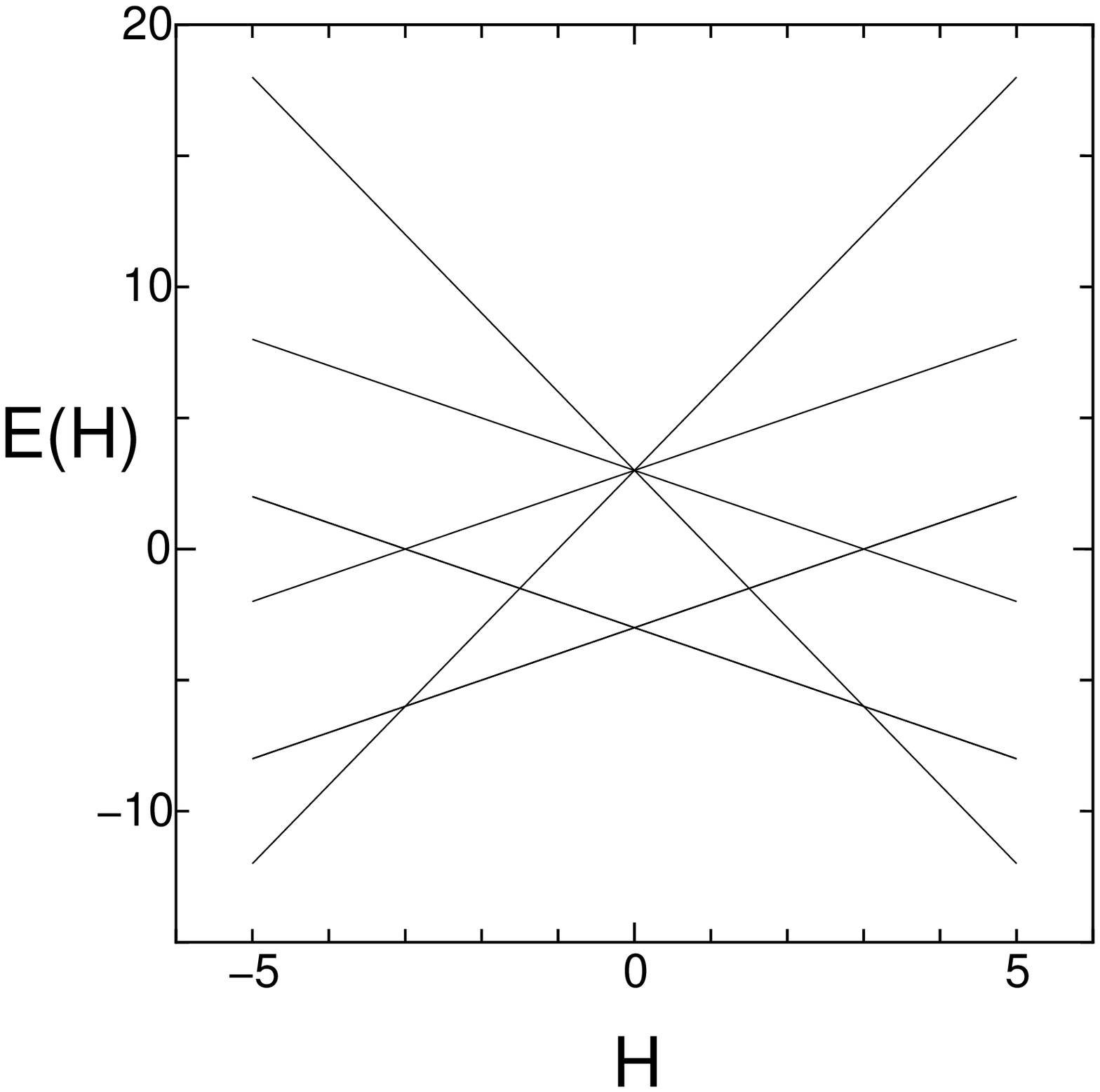} &
\epsfxsize=6.0cm \epsfysize=6.0cm \epsfbox{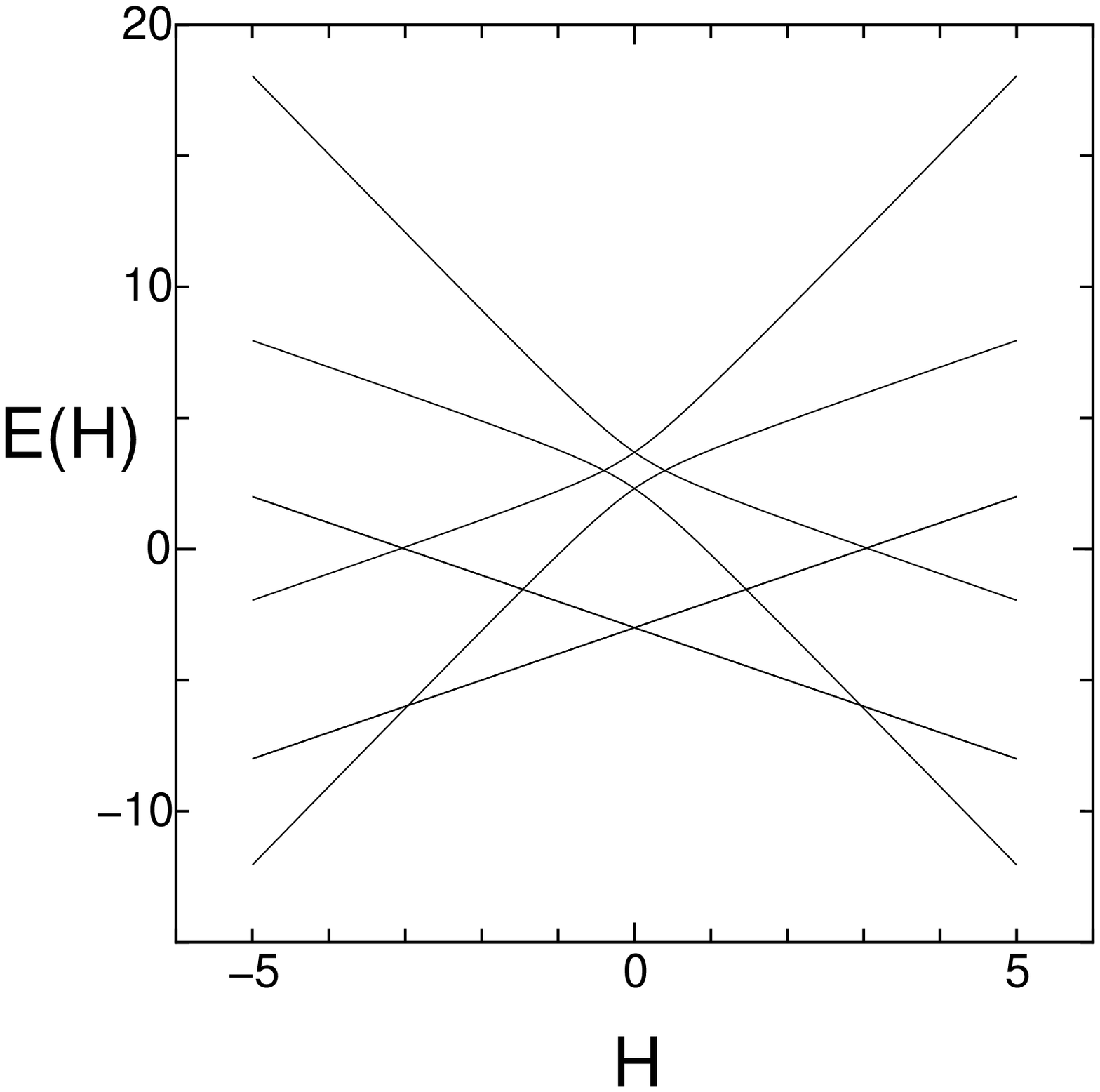}
\end{array}
\nonumber
\end{eqnarray}
\caption{(a) Energy structure of the pure Heisenberg model of $N=3$,
and (b) of the Model I with $\alpha=0.2$}
\label{fig12}
\end{figure}

Here we study an antiferromagnetic Heisenberg chain for ${\cal H}_0$ 
\beq
{\cal H}_0=\sum_{\langle ij\rangle}J_{ij}\mbold{S}_i\cdot \mbold{S}_j
\eeq
with various types of perturbations ${\cal H}_1$ of the form
\beq
{\cal H}_1=\sum_{ij}\alpha_{ij} S_i^zS_j^x.
\label{ZX}
\eeq
First we study a simple case of three spins ($N=3$), and more complicated cases will be
studied later.
The states of the pure Heisenberg model $(\alpha_{ij}=0)$ are classified by the total spin $S$,
that is, the four states of $S=3/2$ and two sets of two states of $S=1/2$ for the present 
case of $N=3$. 
In Fig. \ref{fig12}(a), we show the energy level structure of the system 
for $\alpha=0.0$ as a reference.
Here we see six lines but the levels crossing at low energy consist of
two degenerate states. 
That is, there are four states of $S=1/2$.
This degeneracy for the $S=1/2$ states appears due to the rotational symmetry. 

When we apply the perturbation (\ref{ZX}), the total spin is not a good quantum
number. However, it gives still good description of the states as far as the 
perturbation is small. Thus hereafter we also use $S$ to denote the states.

Here we study the following five categories of the perturbation:

\vspace{0.5cm}
\begin{center}

\Large 

\begin{tabular}{|c|c|c|}
\hline
 Model & Rotational symmetry & Reflection symmetry\\
\hline
I & $\bigcirc \quad \alpha_{12}= \alpha_{23}\cdots = \alpha_{N1}=\alpha $
& $\bigcirc$  \  $\quad \alpha_{ij}=\alpha_{ji}$\\
\hline
II & $\times$ & $\bigcirc$ \ $ \quad \alpha_{12}=\alpha_{21}=\alpha,\quad {\rm others =0}$\\
\hline
III & $\bigcirc$ & $\times$  \ $\quad \alpha_{ij}=-\alpha_{ji}=\alpha$\\
\hline
IV  & $\times$ & $\times$ \ $ \quad \alpha_{12}=-\alpha_{21}=\alpha,\quad {\rm others=0}$\\
\hline
V  & $\times$ & $\times$ \ $ \quad \alpha_{12}\ne\alpha_{21},\quad {\rm others=0}$\\
\hline
\end{tabular}
\vspace{0.3cm}

\normalsize

Table 1 Types of perturbations 
\vspace{0.5cm}

\end{center}

\subsection {Model I}

In Fig. \ref{fig12}(b) we show the energy level structure for Model I with $\alpha=0.2$.
There we find that the hybridization does not occur at lower crossing point
while it occurs at higher crossing point.
In this model the perturbation is symmetric 
for both the rotation and reflection, and
does not affect the symmetry of the wavefunctions. Here we have 
\beq\begin{array}{ll}
{\cal H}_1|{3/2}\rangle=&|{3/2}\rangle \\
{\cal H}_1|{1/2}\rangle=&|{1/2}\rangle,
\end{array}
\eeq
where $|{3/2}\rangle$, and $|{1/2}\rangle$ denote the Hilbert space 
of the wave functions with 
$S=3/2$ and $1/2$, respectively. In particular,
in the space of $S=1/2$ the symmetric operation ${\cal H}_1$ 
does not have matrix elements
\beq
{\cal H}_1|{1/2}\rangle=0.
\eeq
Therefore the degenerate crossing structure of $S=1/2$ states in Fig.\ref{fig12}(a)
is preserved. On the other hand, in the space of $S=3/2$,  ${\cal H}_1$ has
matrix elements between $M=3/2$ and $1/2$, and between $M=-3/2$ and $-1/2$,
where $M$ is the $z$ component of the magnetization.
The absence of matrix elements between $M=1/2$ and $-1/2$  
is due to an peculiar interference of operations of this type of ${\cal H}_1$
(see Appendix).
Thus, avoided level crossing structures are formed only between the states 
$M=3/2$ and $1/2$ and between $M=-3/2$ and $-1/2$.  

Thus if we consider the antiferromagnetic model of the type of Model I, 
the adiabatic transition can not occur between the $M=1/2$ and $-1/2$ states
as we expected from the single spin model. 

In the present paper we will not mention for the ferromagnetic model in detail.
However, it should be noted that the property of the ferromagnetic case is 
simply obtained by reversing the energy axis. 
In the present model, adiabatic transitions occur
between $M=-3/2$ and $-1/2$, and  $M=1/2$ and $3/2$.

\subsection{Model II}
\begin{figure}
\begin{eqnarray}
\begin{array}{ll}
\epsfxsize=6.0cm \epsfysize=6.0cm \epsfbox{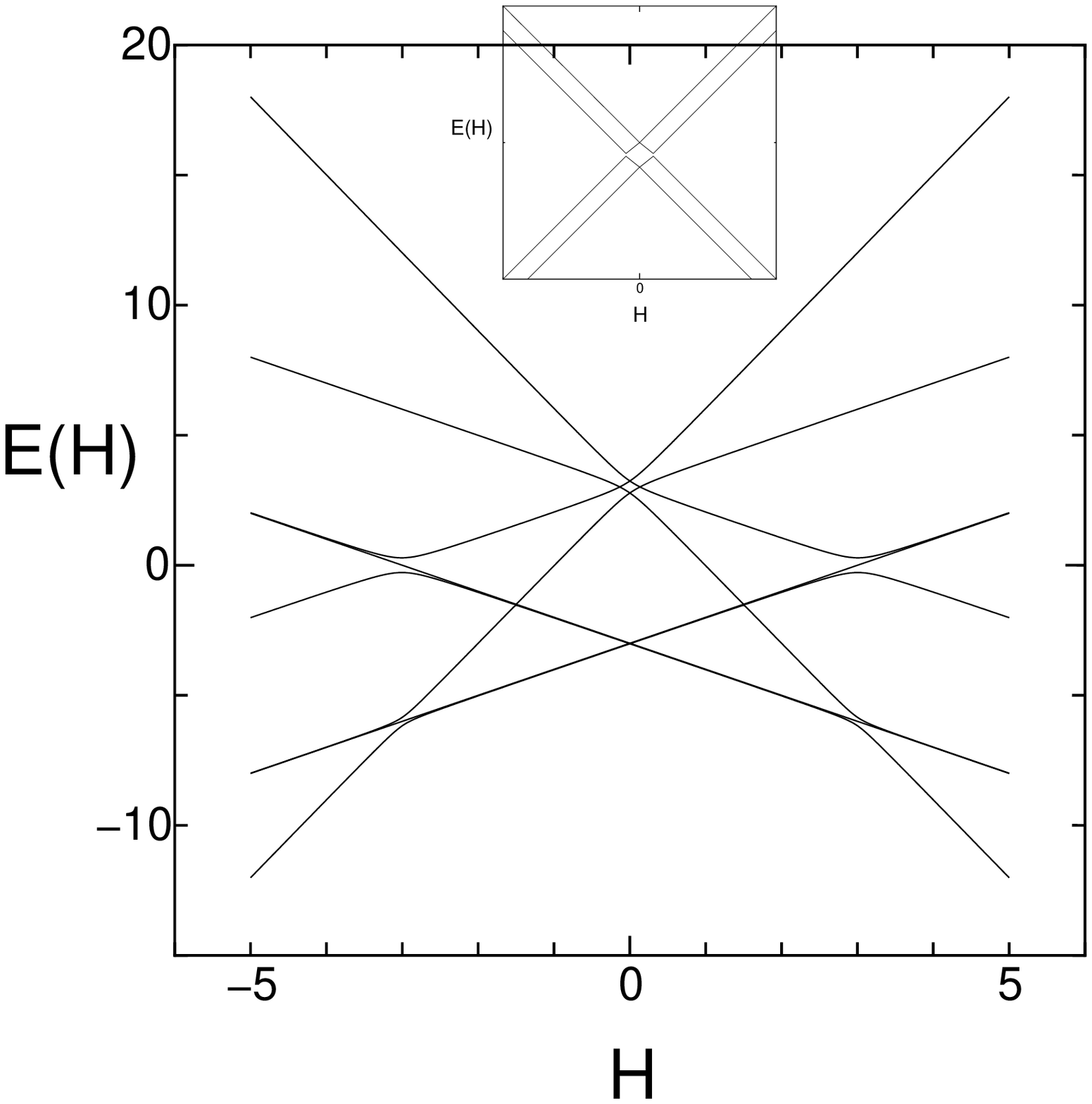} &
\epsfxsize=5.0cm \epsfysize=5.0cm \epsfbox{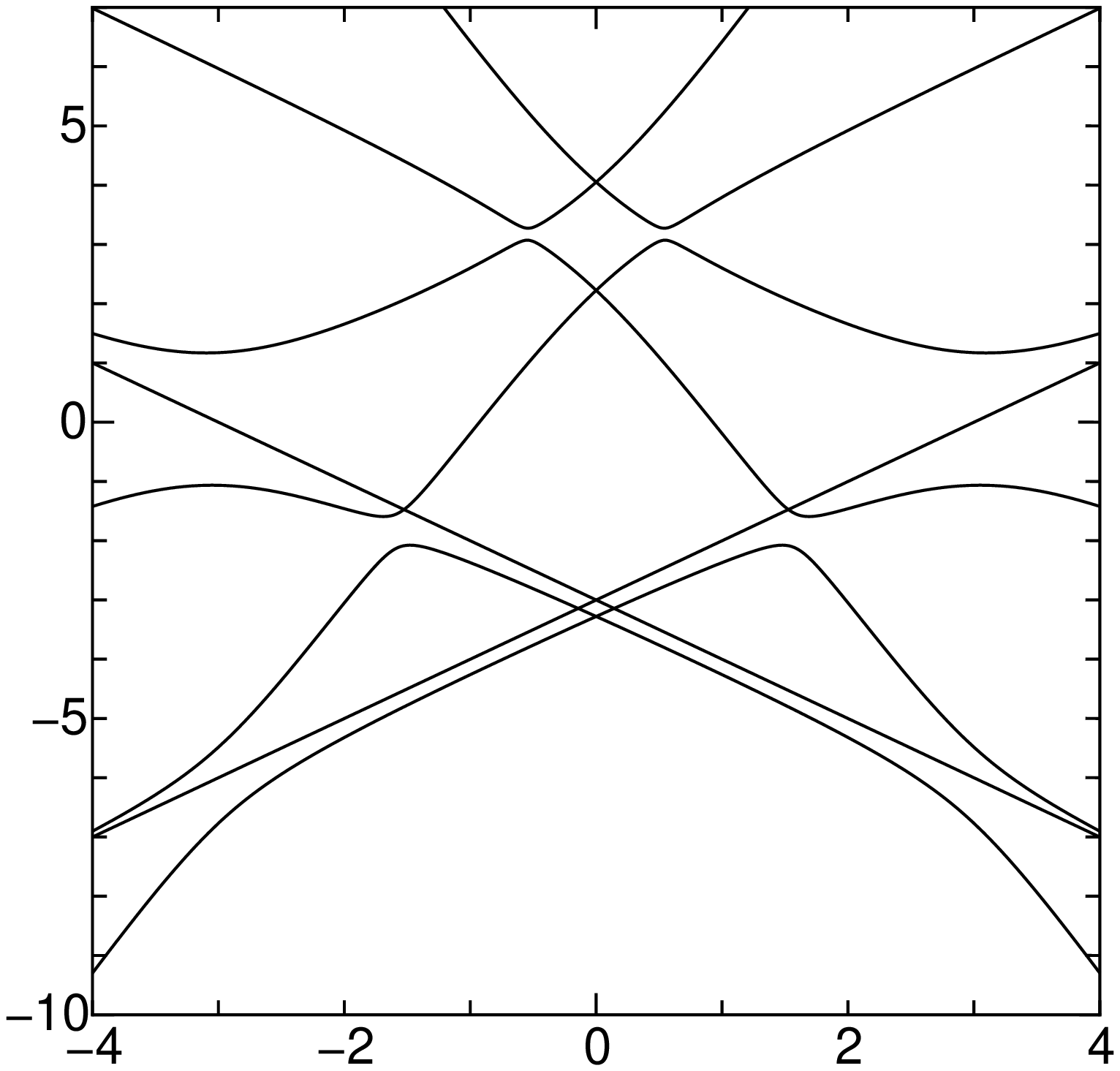}
\end{array}
\nonumber
\end{eqnarray}
\caption{(a) Energy structure of the Model II of $N=3$ with $\alpha=0.2$
(b) with $\alpha=0.8$}
\label{model2}
\end{figure}

Model II of $\alpha=0.2$ shows a similar structure to that 
in Fig. \ref{fig12}(b) as shown in Fig. \ref{model2}(a). 
However due to the lack of the rotational 
symmetry, it shows
more complicated structure as shown in Fig. \ref{model2}(b), where we use 
$\alpha=0.8$ to emphasize the crossing structure.
This model has the reflection symmetry with respect to 
the exchange ${\cal R}_{12}$ of the sites 1 and 2.
Thus the states are classified as
\beq\begin{array}{cl}
{\cal R}_{12}|+\rangle = &+|+\rangle \\
{\cal R}_{12}|-\rangle = &-|-\rangle, 
\end{array}
\eeq
where $|+\rangle$ and $|-\rangle$ denote the symmetric and antisymmetric states
with respect to the exchange ${\cal R}_{12}$.
The four states in $S=1/2$ are separated into sets of $|+\rangle$ and $|-\rangle$.
The antisymmetric states $|-\rangle$ do not contain all up state or all down state,
and are kept in strict $S=1/2$ space. Thus they are not
affected by ${\cal H}_1$ and degenerate at $H=0$.
These two states of $+\rangle$ in $S=1/2$ contain a little components of the 
all up and down state and no more are in $S=1/2$ space strictly. However
the two states of $|+\rangle$ also must degenerate at $H=0$ 
because of the time-reversal symmetry. 
In this model, the adiabatic transition
can not occur between the $M=1/2$ and $-1/2$ states when the field crosses zero
similarly to the previous model. 
The state of $S=3/2$ also forms complicated avoided structure.

\begin{figure}
\begin{eqnarray}
\begin{array}{ll}
\epsfxsize=5.0cm \epsfysize=5.0cm \epsfbox{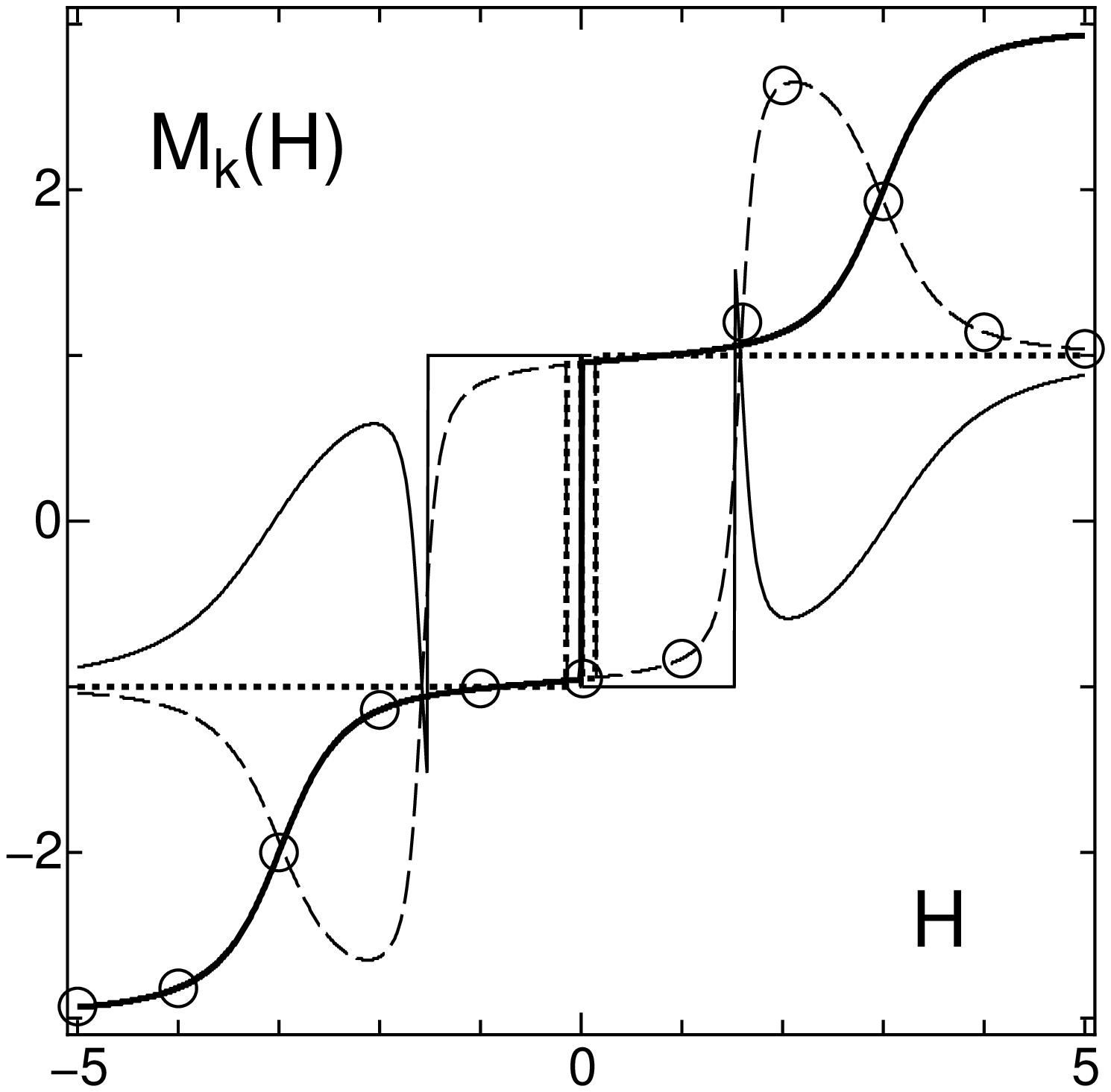} 
\epsfxsize=5.0cm \epsfysize=5.0cm \epsfbox{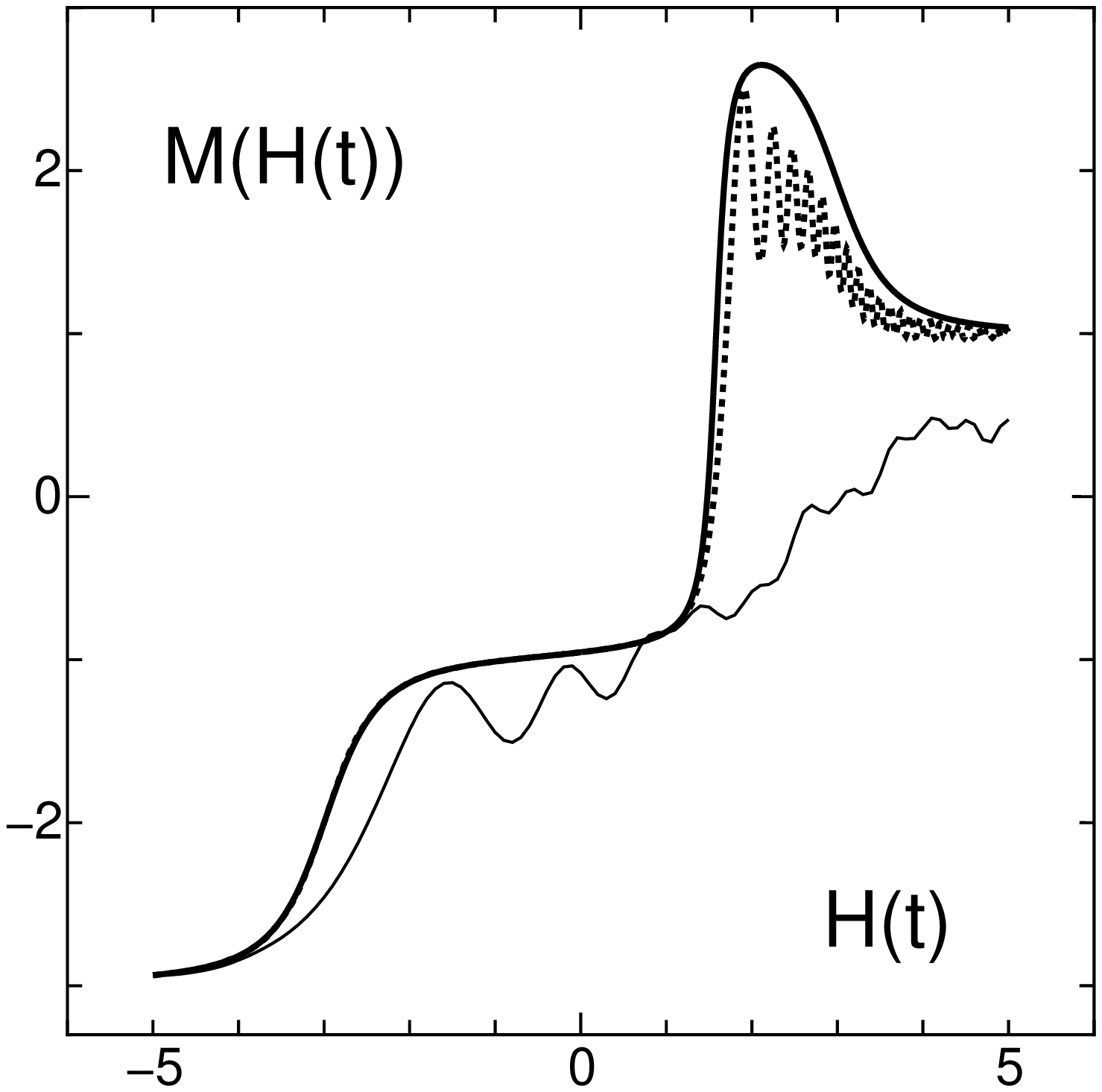} 
\end{array}
\nonumber
\end{eqnarray}
\caption{(a) Adiabatic magnetization process of the Model II with $\alpha=0.8$,
(b) Magnetization process with sweeping field $v=0.02,0.2$ and 2.0, shown
by the solid line, dotted-line and thin solid line, respectively.}
\label{model2-mag}
\end{figure}

In Fig. \ref{model2-mag}(a), we show the magnetization of the states of 
four low energies as functions of the field for the case of $\alpha=0.8$. 
The bold solid line,
bold dotted-line, dashed-line and thin solid line denote the magnetization 
of the levels 1,2,3 and 4, respectively.
When the levels cross the order of the lines changes. Therefore
the types of lines change discontinuously.
However, the adiabatic motion of magnetization is given by 
smooth continuation of a level. The adiabatic magnetization 
starting with the ground state at $H=-5$ is shown by big circles.
It shows a strange field dependence. That is, around $H\simeq 2$ the magnetization
has a peak, where the coefficient of the all up state becomes large.

When we sweep the field,
\beq
H(t)=-H_0+vt,
\eeq
the evolution of the state is expressed by
\beq
|\Psi(t)\rangle= \exp_{\rm t}
\left(-i\hbar\int_{t}^{t+\Delta t}{\cal H}(s)ds\right) |\Psi(t)\rangle,
\quad {\cal H}(t)={\cal H}-H(t)\sum_{i}S_i^z,
\label{QEV}
\eeq   
where $\exp_{\rm t}$ means the time ordered exponential.
In Fig. \ref{model2-mag}(b), the magnetization process with sweeping field with
velocity $v=0.02,0.2$ and 2.0, are shown. In the case of $v=0.02$ the magnetization 
in the time dependent process reproduces well the adiabatic one.  

Here, it should be noted that the magnetization process shows a non-monotonic
dependence on the field, which violates the thermodynamical stability. That is,
the Zeeman energy (\ref{Zeeman}) decreases when the field increases.
Namely, if we increase the magnetic field the magnetic field
receives some energy but not gives the energy as in the normal cases.
This thermodynamically unstable behavior comes from the fact that
this adiabatic state is no more the ground state.
In the ground state the Zeeman energy must increase monotonically with $H$.
In the present case,
the ground state for $H<0$ adiabatically becomes an excited state 
after the crossing point due to the crossing.
It would be interesting to find phenomenon corresponding to the present
observation and to make use this peculiar behavior of the magnetization.
  
\subsection{Model III}

\begin{figure}
\vspace*{-1cm}
\begin{eqnarray}
\begin{array}{ll}
\epsfxsize=6.0cm \epsfysize=6.0cm \epsfbox{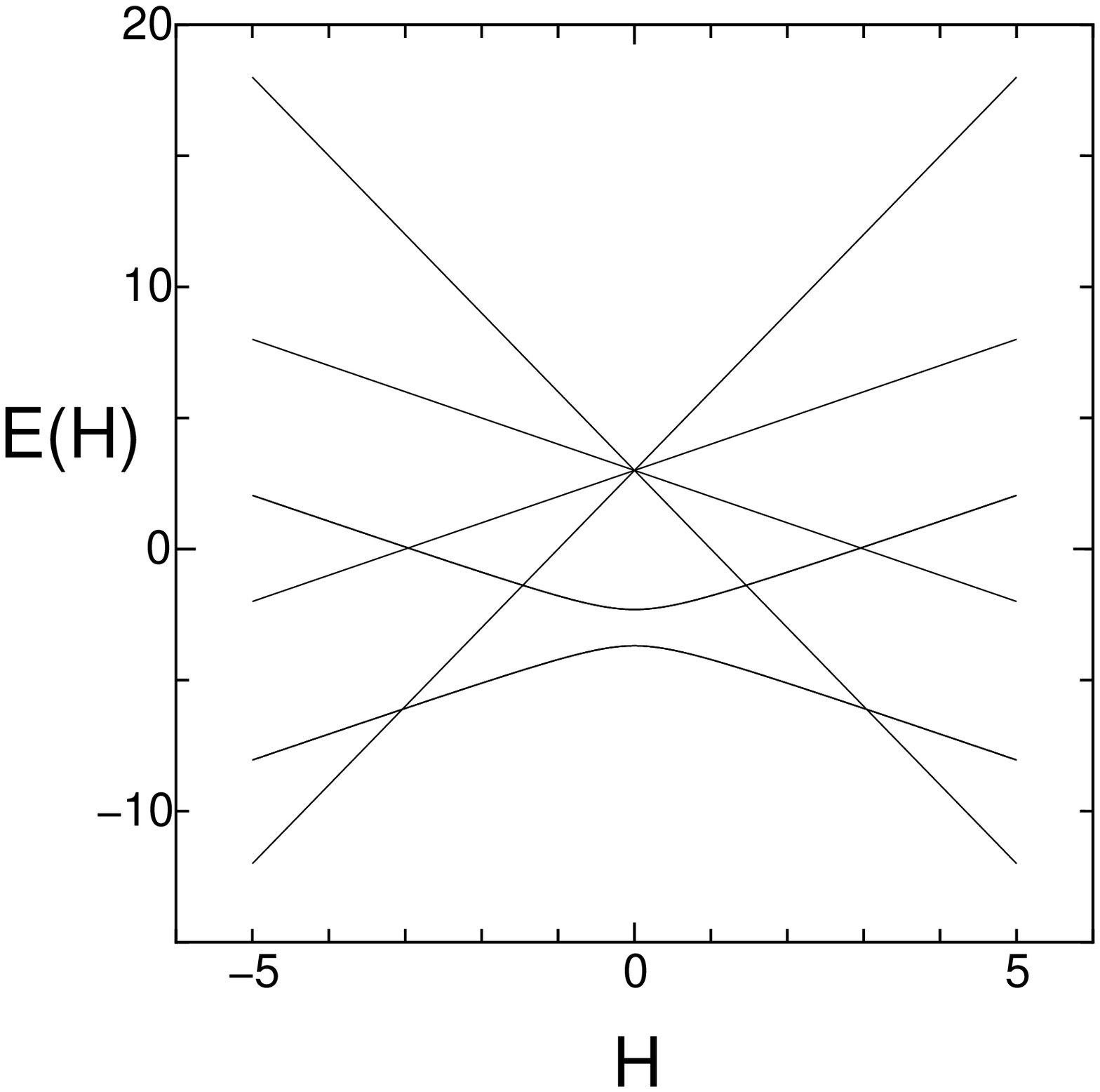}
\epsfxsize=5.0cm \epsfysize=5.0cm \epsfbox{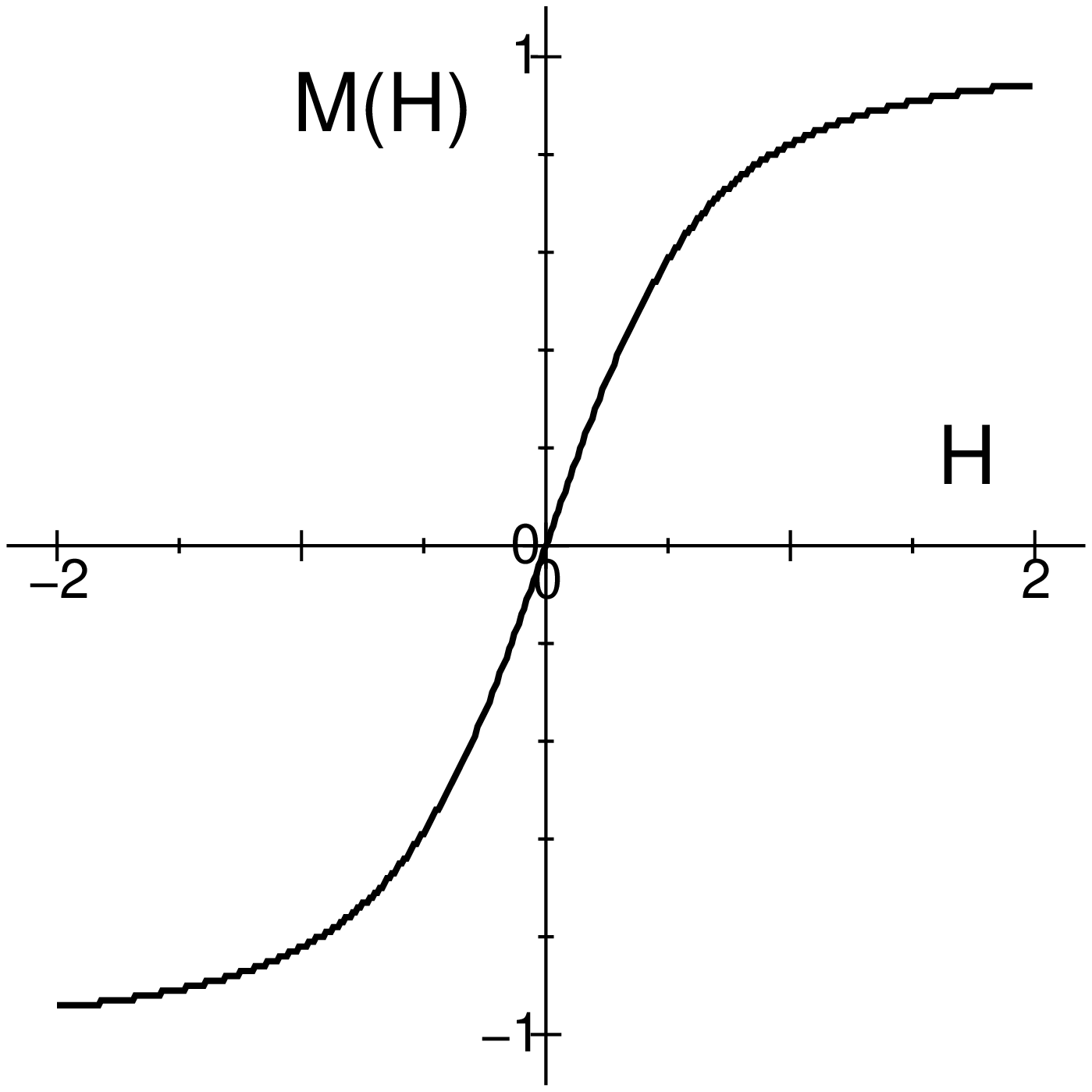}
\end{array}
\nonumber
\end{eqnarray}
\caption{ (a)
Energy structure and (b) ground state magnetization, 
of the Model III of $N=3$ with $\alpha=0.2$}
\label{model3}
\end{figure}
In Fig.  \ref{model3}(a), 
we show the energy level structure for Model III with $\alpha=0.2$. 
Using a similar argument for Model I, 
\beq
{\cal H}_1|3/2\rangle=0,
\eeq
and thus
we find that the structure of $S=3/2$ states is preserved.
On the other hand, ${\cal H}_1$ can change the magnetization in
the space of $S=1/2$. Here the states of $S=1/2$ with different magnetizations
$M=\pm 1/2$ are hybridized and form two sets of avoided level crossing
structure. 
The magnetization process of the ground state is shown in Fig.  \ref{model3}(b).
They are completely degenerate due to the rotational symmetry.
The adiabatic transition 
between the different sets of avoided cross structure is prohibited.
Let us take a ground state $|a\rangle$ at $H=-H_0$ in 
one set of avoided cross structure as an initial state. 
We have another ground state $|b\rangle$ in the other set 
which is orthogonal to $|a\rangle$.
The orthogonality does not change by the sweeping operation because of the symmetry.
For example if we begin with a state whose eigenvalue is $e^{4i\pi/3}$ for
one third rotation, the state with the eigenvalue $e^{8i\pi/3}$
is not mixed at all. 
Thus we can consider the time evolution for both states separately.
In each branch, the nonadiabatic transition probability is exactly 
given by the equation (\ref{LZS}), and thus the total change of the magnetization
for any combination is given by the LZS formula.

\subsection {Model IV}

\begin{figure}
\begin{eqnarray}
\begin{array}{ll}
\epsfxsize=6.0cm \epsfysize=6.0cm \epsfbox{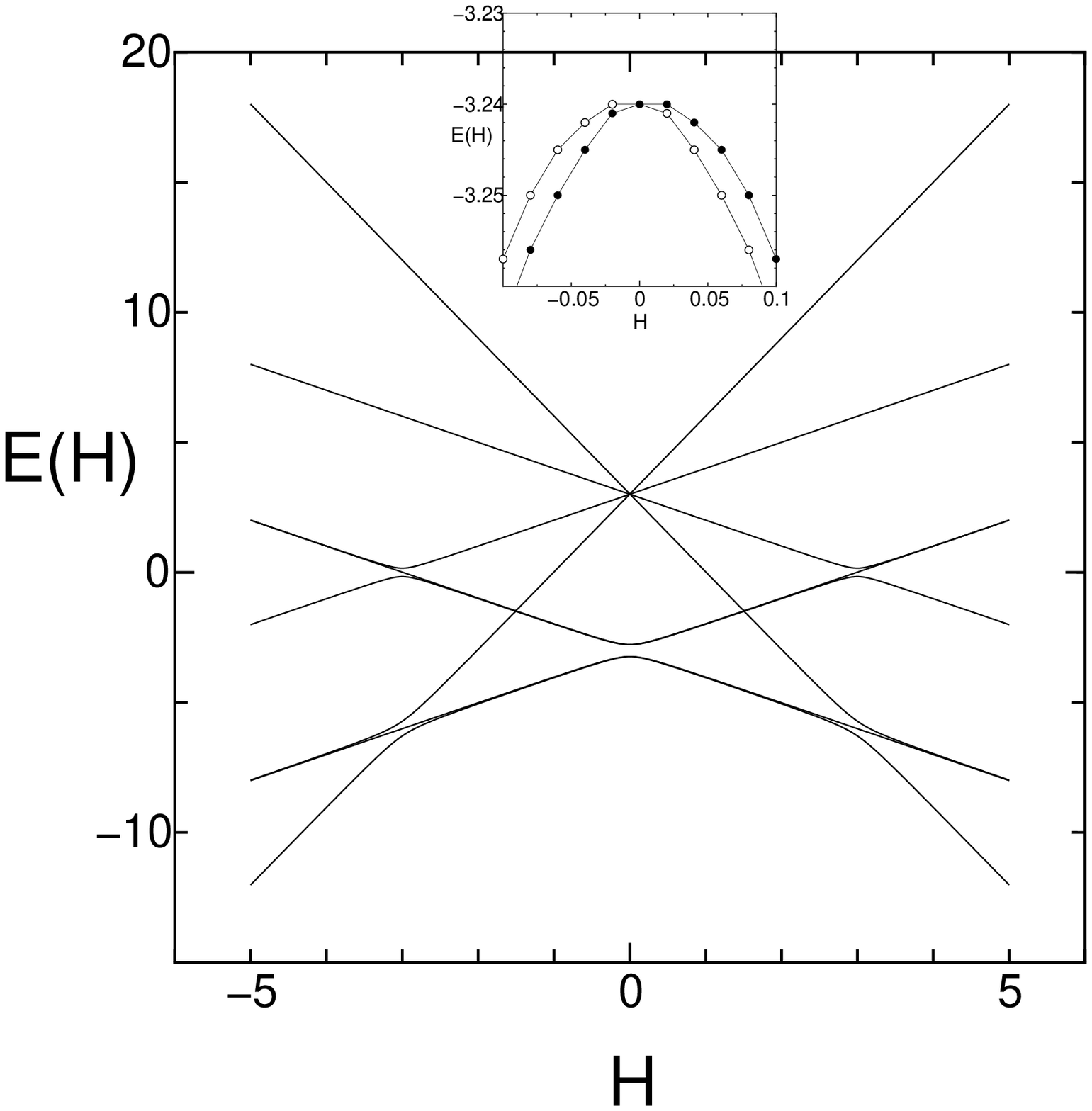} &
\epsfxsize=5.0cm \epsfysize=5.0cm \epsfbox{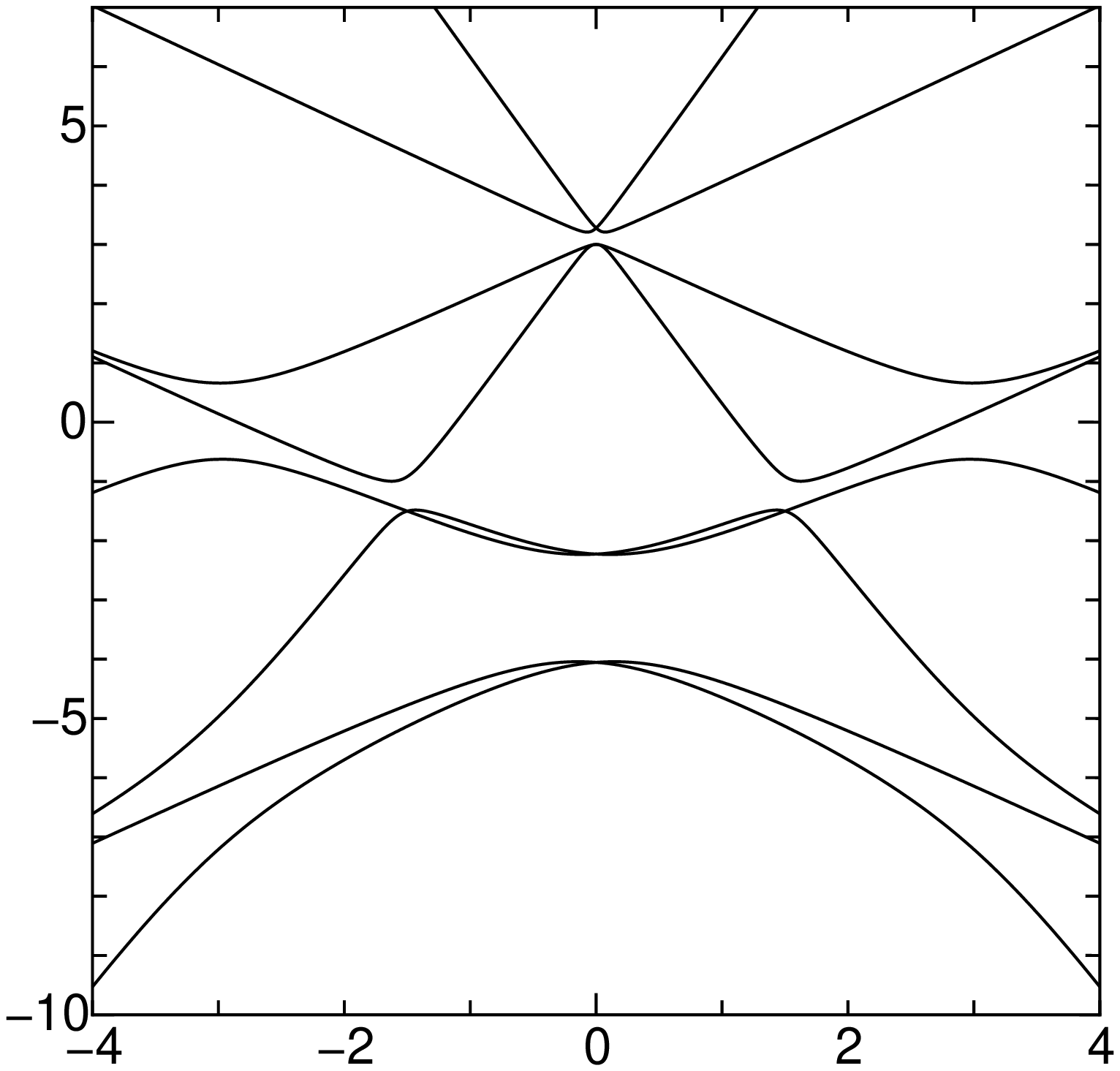}
\end{array}
\nonumber
\end{eqnarray}
\caption{(a) Energy structure of the Model IV with $\alpha=0.2$,
(b)  with $\alpha=0.8$}
\label{model4}
\end{figure}
In Fig.  \ref{model4}(a), we show the energy level structure for Model IV with $\alpha=0.2$.
There we find a similar structure to that of Model III.
However, because of the lack of the symmetry, the four levels of 
low energies are no more degenerate  and if we look the state carefully
they have  different field dependences (see the inset).
In order to emphasize the structure, the energy structure of $\alpha=0.8$ is also 
shown in  Fig.  \ref{model4}(b). 
Although this model has no geometrical symmetry, the model has a peculiar
symmetry due to the special interference in the triangle lattice.
That is, the all up state $|+++\rangle$ can not reach to the all down state $|---\rangle$.
Thus the states is classified into the following two categories
\beq
\begin{array}{ll}
|\rm{A}\rangle=& a|+++\rangle + b|+\rangle_{\rm o} + c|-\rangle_{\rm e}\\   
|\rm{B}\rangle=& a'|---\rangle + b'|-\rangle_{\rm o} + c'|+\rangle_{\rm e},  
\end{array}
\eeq
where $|+\rangle_{\rm e}$ is a symmetric state of $M=1/2$ with respect to 
the exchange of the sites 1 and 2, and  $|-\rangle_{\rm o}$ is the
antisymmetric state of $M=-1/2$, and so on.
In the inset of \ref{model4}(a), the open circles denote the level of $|\rm{A}\rangle$
and the closed circles $|\rm{B}\rangle$.  
Avoided crossing structure is formed in each category. 
Thus there are no mixing between different sets of the avoided crossing structure.
Because of the Kramers theorem, they cross at $H=0$ and recover the degeneracy.
For any combination of the initial populations of the two sets,
the nonadiabatic transition probability in each set is given by
(\ref{LZS}), and the change of the total magnetization follows the LZS mechanism.
In this model, if we sweep the field in a wide range, e.g., from $-4$ to 4,
the magnetization adiabatically changes from $M=-3/2$ to $M=1/2$, but not
to $3/2$. This asymmetry of the adiabatic change is one of the 
characteristics of the present model. 

Up to the model IV, the system has some symmetry with respect to the
exchange of spin 1 and 2. Now we try to remove this symmetry. 
Now we put only $\alpha S_1^xS_2^z$, but                  
there we found essentially the same structure to that of Model IV.
We found that there is still some symmetry which causes the 
decoupling of the space into $2+2$.
Even if we add a term $\beta S_2^yS_3^z$ in order to reduce possible 
symmetry, we find that there still exists two sets of states.
These separations are due to rather special interference on the triangle structure. 

\subsection{Model V}

\begin{wrapfigure}{r}{6.6cm}
\vspace*{0cm}
\begin{eqnarray}
\begin{array}{ll}
\epsfxsize=5.0cm \epsfysize=5.0cm \epsfbox{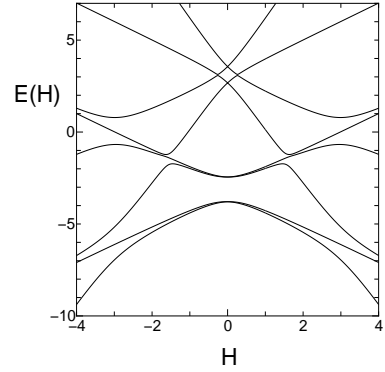}
\end{array}
\nonumber
\end{eqnarray}
\caption{ Energy structure of the Model V of $N=3$ with $\alpha=0.2$
$\alpha_{21}=-1.0$.}
\label{model5e}
\end{wrapfigure}

\begin{figure}
\begin{eqnarray}
\begin{array}{ll}
\epsfxsize=5.0cm \epsfysize=5.0cm \epsfbox{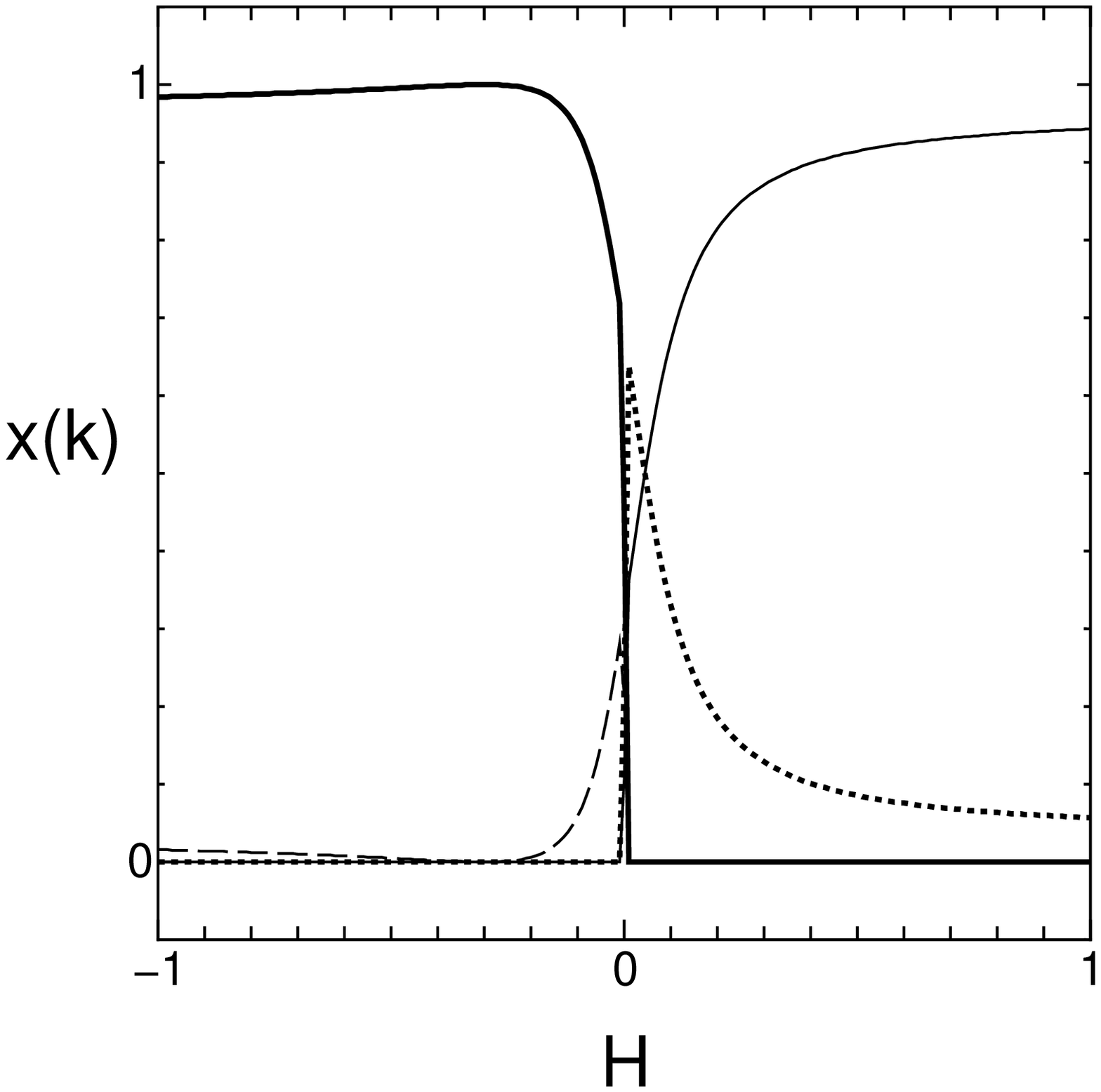} &
\epsfxsize=5.0cm \epsfysize=5.0cm \epsfbox{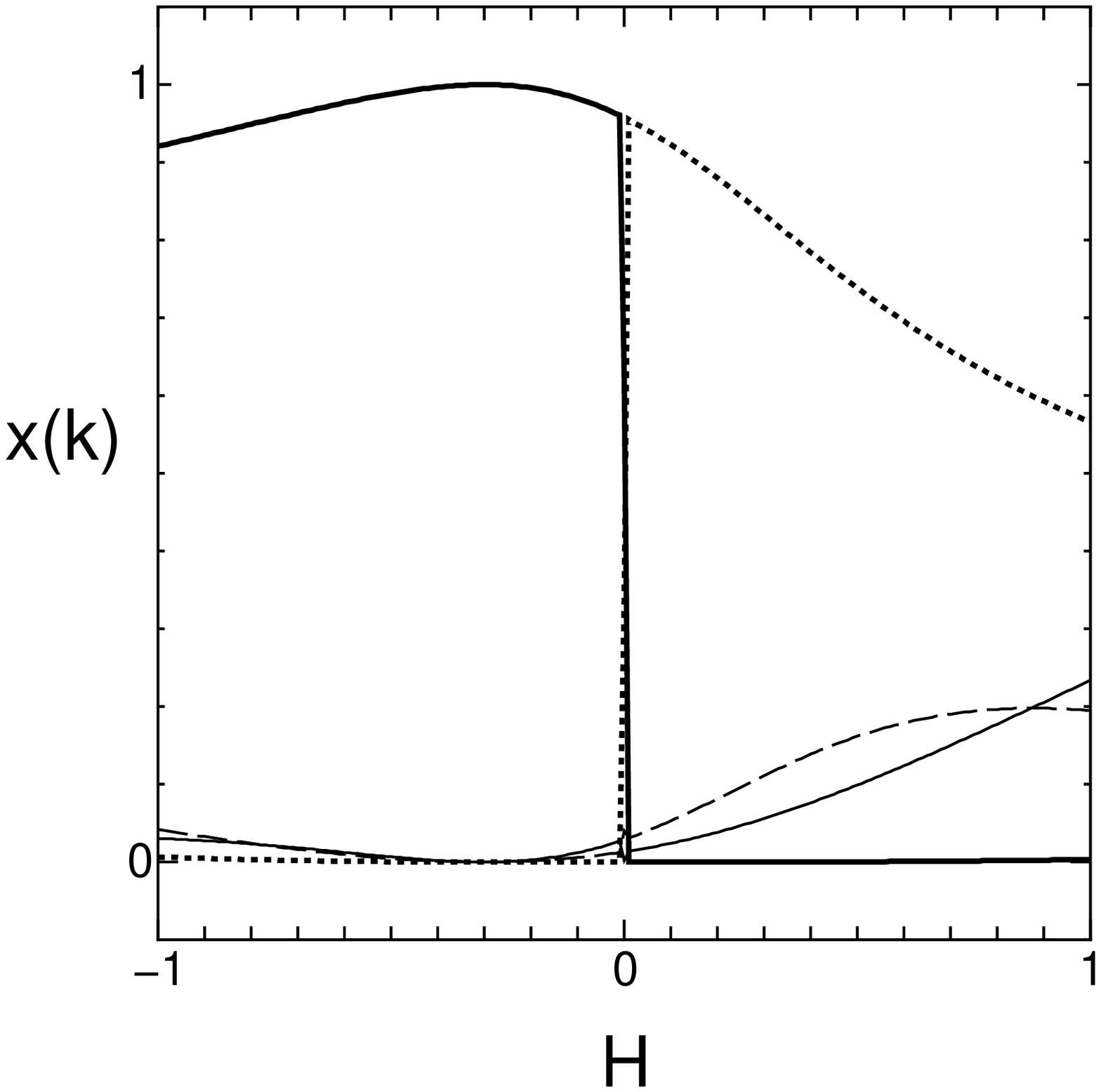}
\end{array}
\nonumber
\end{eqnarray}
\caption{(a) Overlap $x(k)$ for $\alpha_{12}=0.2$ and $\alpha_{21}=0$ and 
(b) for $\alpha_{12}=0.2$ and $\alpha_{21}=-1.0$} 
\label{model5x}
\end{figure}

Finally we put $\alpha_{12}=0.2$ and $\alpha_{21}=-1.0$.
In Fig. \ref{model5e}, we show the energy level structure for this case,
which is similar to the previous one.
But in this model we found that there is no separation into $2+2$ subspaces
any more. 
The overlaps of the ground state at $H=-0.3$ ($|G(H=-0.3)\rangle$) 
and the eigenstates at $H$ ($|\phi_k(H)\rangle$ at $H$) of four lowest
eigenvalues are measured by the quantity
\beq
x(k)=|\langle G(H=-0.3)|\phi_k(H)\rangle|^2, \quad
 k=1,\cdots,4,
\label{OVLP}
\eeq 
as shown in Figs.\ref{model5x}.
In Fig. \ref{model5x}(a), we show the overlaps in a case of 
$\alpha_{12}=0.2$ and $\alpha_{21}=0$ where
the separation into $2+2$ occurs, and in Fig. \ref{model5x}(b) 
the present case ($\alpha_{12}=0.2$ and $\alpha_{21}=-1.0$).
Here the data for $k=1,\cdots,4$  
are shown by  bold-solid, bold-dotted, dashed and thin-solid lines,
respectively.  While the set of states of $k=1$ and 3 and that of $k=2$ and 4 
do not
mix at all in Fig. \ref{model5x}(a), all the four states have some overlap 
with  the state $|G(H=-0.3)\rangle$ in Fig. \ref{model5x}(b). 
The bold line in Fig. \ref{model5x}(b) shows very small but nonzero values
at $H>0$.
The wave functions consist of all the states
\beq
|\Psi(t)\rangle =
 a|+++\rangle + b|+\rangle_{\rm o} + c|-\rangle_{\rm e}
+ a'|---\rangle + b'|-\rangle_{\rm o} + c'|+\rangle_{\rm e}.
\eeq
Thus, the simple LZS transition
mechanism does not work in this model. Dynamical property of this model
will be studied in the next section. 

\subsection{Summary}

Here we study the simple case of $N=3$, but we can learn general properties 
for higher spins. That is, although the time-reversal symmetry prohibits the
tunneling between the Kramers doublets at $H=0$, it does not means 
the absence of the adiabatic change of magnetization when the field swept at $H=0$.  
If the system has additional degree of freedom and more than two 
 states are degenerate, e.g. 4-fold degeneracy, 
at $H=0$ in the unperturbed system ${\cal H}_0$, 
the avoided level crossing structure is formed with respect to the 
field as in the cases of Model III, IV, and V. 
If the system has additional symmetries, then the avoided level crossing 
structures are classified by the symmetry and they are independent each other,
where the LZS transition mechanism works as in the two-level system.
On the other hand, in the absence of additional symmetry,
the degenerate states forms an irreducible space except $H=0$. There
the simple LZS transition mechanism does not work.
In the next section we study the magnetization change under the sweeping field
in the latter case.

\section{Quasi-Landau-Zener-St\"uckelberg Transition}

\begin{figure}
\begin{center}
\hspace*{0.8cm}
\epsfxsize=6.0cm \epsfysize=6.0cm \epsfbox{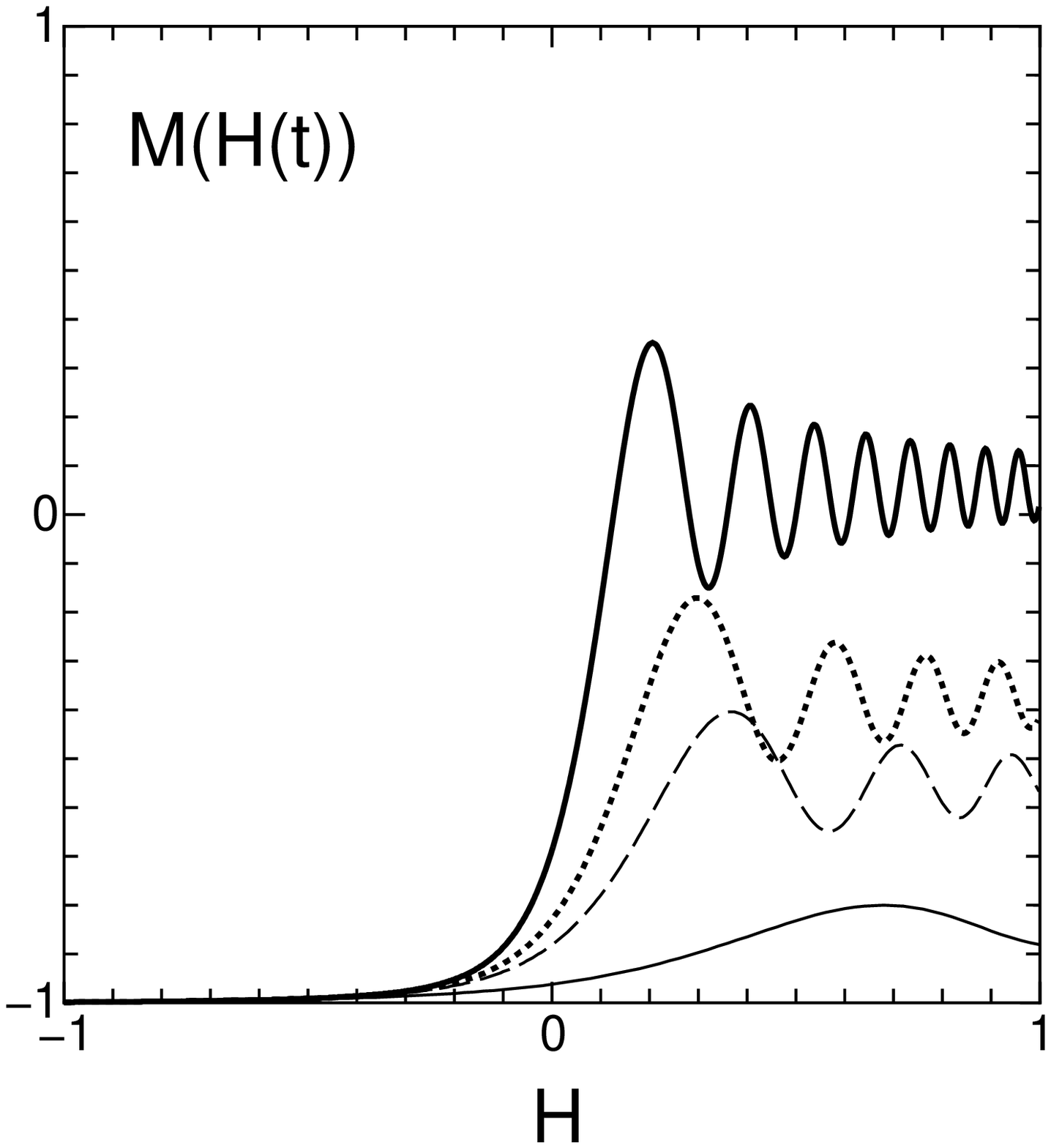} 
\epsfxsize=6.0cm \epsfysize=6.0cm \epsfbox{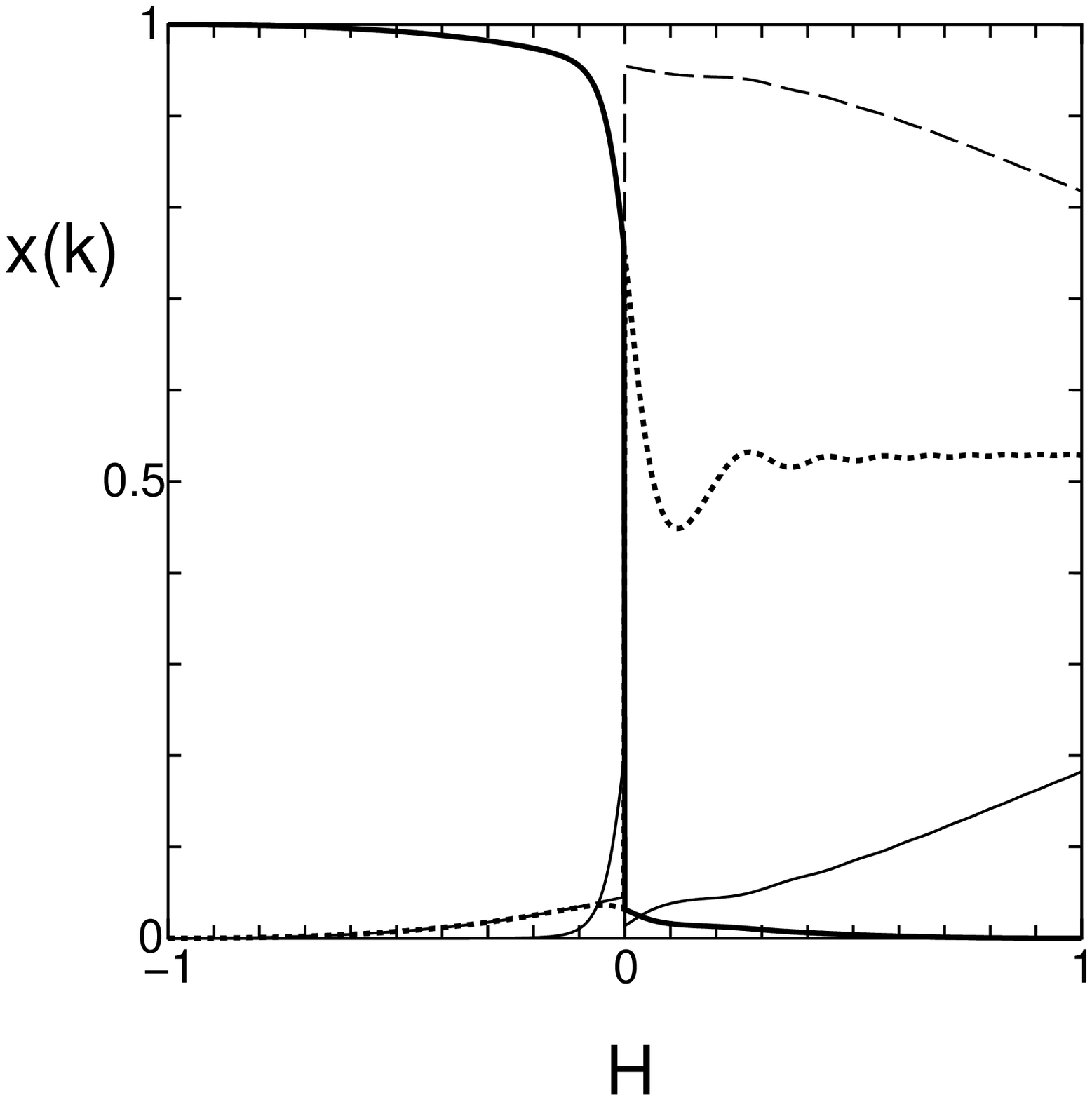} 
\end{center} 
\caption{(a) Magnetization process in the sweeping speed $v=$ 0.04, 0.08,
0.12, and 0.4 from the top to the bottom. $\alpha_{12}=0.02, \alpha_{21} = - 0.1$ 
(b) The overlaps with the
$k$-th adiabatic state $p_k(H)$ for k=1,2,3 and 4 by the
bold solid, bold dotted, dashed and solid lines, respectively. 
 $\alpha_{12}=0.02, \alpha_{21} = - 0.1$, $v=$ 0.04. }    
\label{302-5mx}
\end{figure}
Now we demonstrate the time dependent process with sweeping field 
where $H$ is swept from $-H_0$ to $H_0$.
\beq
|\Psi(t_{\rm f})\rangle=\exp_{\rm t}\left(i\int_{t_{\rm i}}^{t_{\rm f}}{\cal H}(s)ds\right)
|G(H=-H_0)\rangle,
\eeq
where $|G(H)\rangle$ is the ground state for $H$, 
and $H(t_{\rm i})=-H_0$ and  $H(t_{\rm f})=H_0$. 
In Fig. \ref{302-5mx}(a) we show the magnetization for the case of 
$\alpha_{12}=0.02$ and $\alpha_{21}=-0.1$ (Model V).
Here we take small values of $\alpha$'s in order to facilitate the observation 
of the nonadiabatic transition. 
The speed $v$ of the field-sweep is
 0.02, 0.04, 0.08, and 0.2 from the top to the bottom, respectively.
As we saw in the previous section, 
the simple LZS mechanism can not be applied for this model because all the 
four low energy states contribute to this transition.
However we find similar $v$-dependence of the magnetization change to that in the 
LZS mechanism.
In order to study the detail of the transition we investigate  
the population of the $k$-th eigenstate at the field $H$ by
\beq
p_k(H)=|\langle \Psi(t)|\phi_k(H)\rangle|^2, \quad H=H(t).
\eeq
In Fig. \ref{302-5mx}(b), the change of the populations are shown, where we find
the ground state at large negative field is scattered to all the
four state and they change with the field.
This is a characteristic feature of the present model in contrast with the other cases:
e.g., in the case of $\alpha_{21}=0$, where the four states are separated into 
two sets and the ground state at high negative field is scattered 
only to the second and forth states.  
\vspace{0.5cm}

\begin{figure}
\vspace*{-0.2cm}
\begin{center}
\hspace*{0.5cm}
\epsfxsize=6.5cm \epsfysize=6.0cm \epsfbox{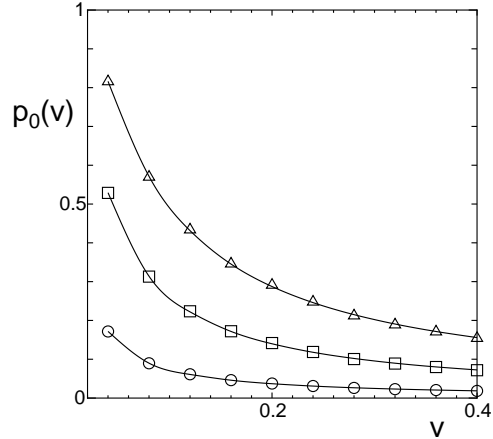} 
\end{center} 
\vspace{-0.2cm}
\caption{ Nonadiabatic transition probabilities
 in the Model V with $\alpha_{12}=-\alpha_{21}/5=0.03$, 0.02, and 0.01,
from the top to the bottom. The solid curves represents the value estimated
by the LZS formula (\ref{LZS}).}
\label{LZSprb}
\end{figure}
However, if we sum up the populations of the first two states, 
\beq
p_0=|\langle \Psi(t_f)|\phi_1(H)\rangle|^2+ |\langle \Psi(t_f)|\phi_2(H)\rangle|^2,
\eeq
the sum almost satisfies the dependence of LZS transition probability (\ref{LZS}) as
shown in Fig. \ref{LZSprb} where    
\beq p_0=1-p.  \eeq
Here we use the energy gap at $H=0$ as $\Delta E$.
The data for the system of $\alpha_{12}=-\alpha_{21}/5=0.03$, 0.02, and 0.01 
are shown by circle, square, and triangular, respectively. 
The lines denote the corresponding dependences due to the relation  (\ref{LZS}).

This is a problem of nonadiabatic transition
of four levels. But the mechanism of the present model 
seems to belong to a different type 
from that of the bow-tie model\cite{BT}.
The present transition may be attributed to some multiple 
crossings\cite{KF,Kb}, and then have some relation with the Brundobler and Elser 
hypothesis\cite{B-E}. 
The mechanism of this quasi-LZS behavior of the sum of the population 
would be an interesting problem in the future. 

\begin{figure}
\begin{eqnarray}
\begin{array}{ll}
\epsfxsize=6.0cm \epsfysize=6.0cm \epsfbox{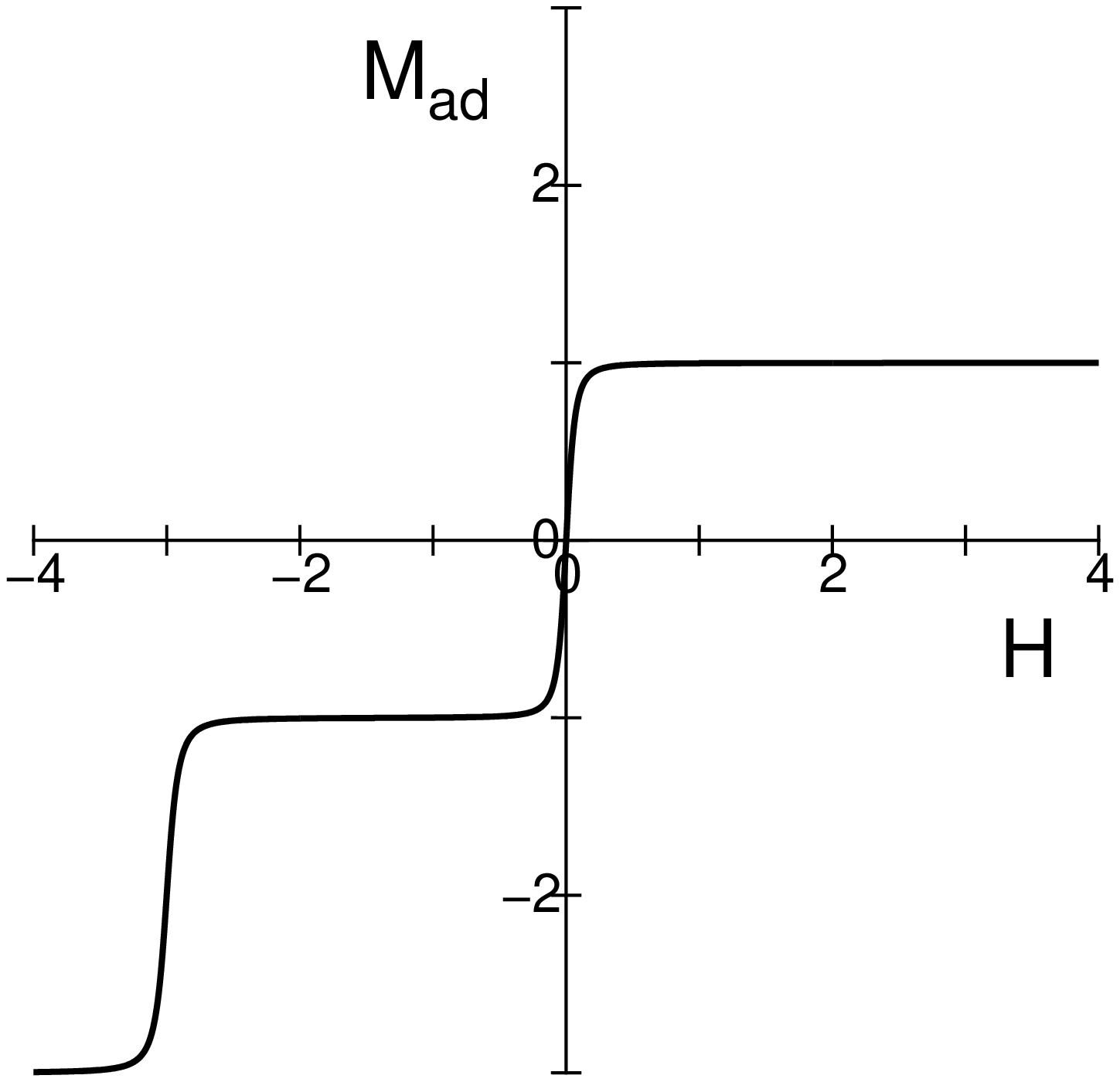} &
\epsfxsize=6.0cm \epsfysize=6.0cm \epsfbox{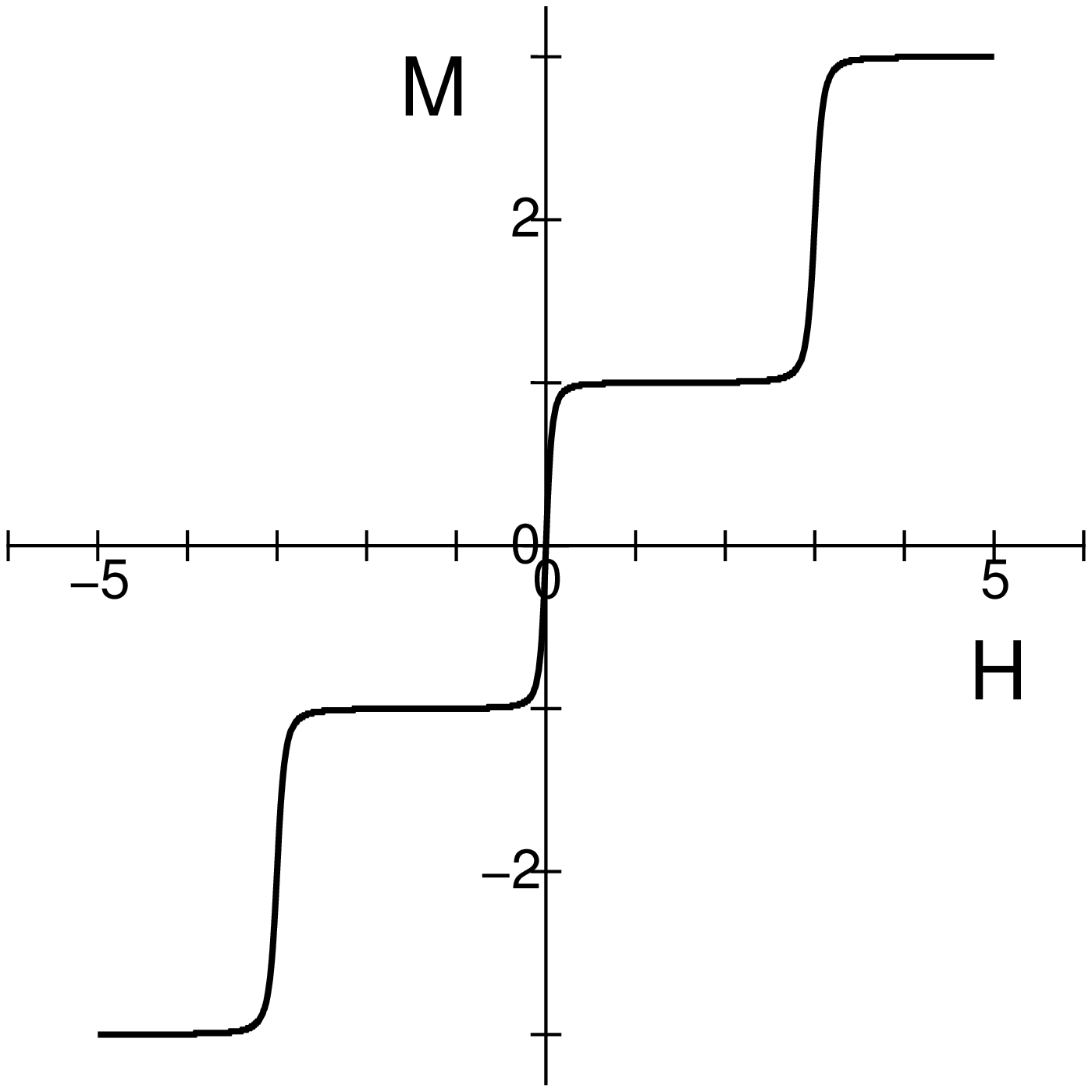}
\end{array}
\nonumber
\end{eqnarray}
\caption{(a) Adiabatic magnetization process ($v=0.002$) and
(b) Magnetization process of the ground state, 
of Model V for $\alpha_{12}=0.02$ and $\alpha_{21}=-0.1$}  
\label{figs302mag}
\end{figure}

From a view point of the present study, 
the resonant tunneling from $-1/2$ to $1/2$ at $H=0$ 
in molecular magnets with half-inter spin such as V$_{15}$ 
does not contradict with general theory of the symmetry.
Here it would be interesting to point out the following fact.
If the field is swept slowly from a large negative value where
the magnetization is $-3/2$, the magnetization change adiabatically 
as $-3/2 \rightarrow -1/2 \rightarrow 1/2$, but not to $3/2$ as shown 
in Fig. \ref{figs302mag}(a),
although the ground state magnetization process is, of course, symmetric
(Fig. \ref{figs302mag}(b)). 
We found this behavior exactly in the Model IV. However, in the present model
the wave functions contain both components of $|+++\rangle$ and $|---\rangle$.
Therefore this asymmetry of the adiabatic magnetization seems to occur 
widely in model on the triangle. 
If the system essentially consists of three spins, this asymmetric 
adiabatic process would be
observed in a slow sweeping field at low temperatures. However,
if other spins strongly contribute to the interaction,
the states of $3/2$ could be also degenerate. 
In such cases also 
the transition from $1/2$ to $3/2$ would be allowed. It would be
an interesting problem to study the magnetization change of V$_{15}$
in a slowly sweeping field , i.e., adiabatic change of the field. 
This enables us to determine  whether
only three spins in the middle of the molecule mainly contribute
to the interaction, or the whole 15 spins contribute. 
We hope that some experimental study on this problem will be done, although 
the thermal effects would easily smear out the pure quantum selection rule.

\section{Higher Spin Cases}

\begin{wrapfigure}{r}{6.6cm}
\begin{center}
\hspace*{1.2cm}
\vspace*{-0.5cm}
\epsfxsize=3.7cm \epsfysize=3.7cm \epsfbox{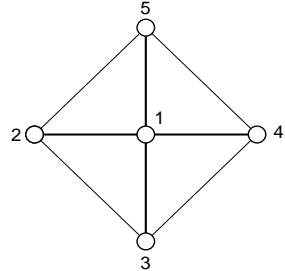} 
\end{center} 
\vspace*{ 0.2cm}
\caption{Lattice of five spins}
\label{spin5}
\end{wrapfigure}
So far we studied magnetic properties the crossings  
which are mainly consists of $S=1/2$ states. 
Here let us study the cases of higher spins.
As an example, we study a system of five spins shown in Fig. \ref{spin5},
where the bold line denotes $J$, and the thin line denotes $J'$.
The perturbation (\ref{ZX}) is set between the sites 1 and 2 
only with $\alpha_{12}=\alpha$.

If we set $J=J'=1$ and $\alpha=0$, the ground state at $H=0$ is 
given by 4-fold degenerate $S=1/2$ states, 
and the energy structure is similar to Fig.\ref{fig12}(a). 
If we put $\alpha=0.2$, the energy structure is similar to Fig. \ref{model4}(a), 
and we find two sets of independent avoided level crossing structure
again. Thus, it is expected that the LZS transition works in a wide range of 
systems when the ground state is in $S=1/2$.

\begin{figure}
\begin{eqnarray}
\begin{array}{ll}
\epsfxsize=6cm \epsfysize=6cm \epsfbox{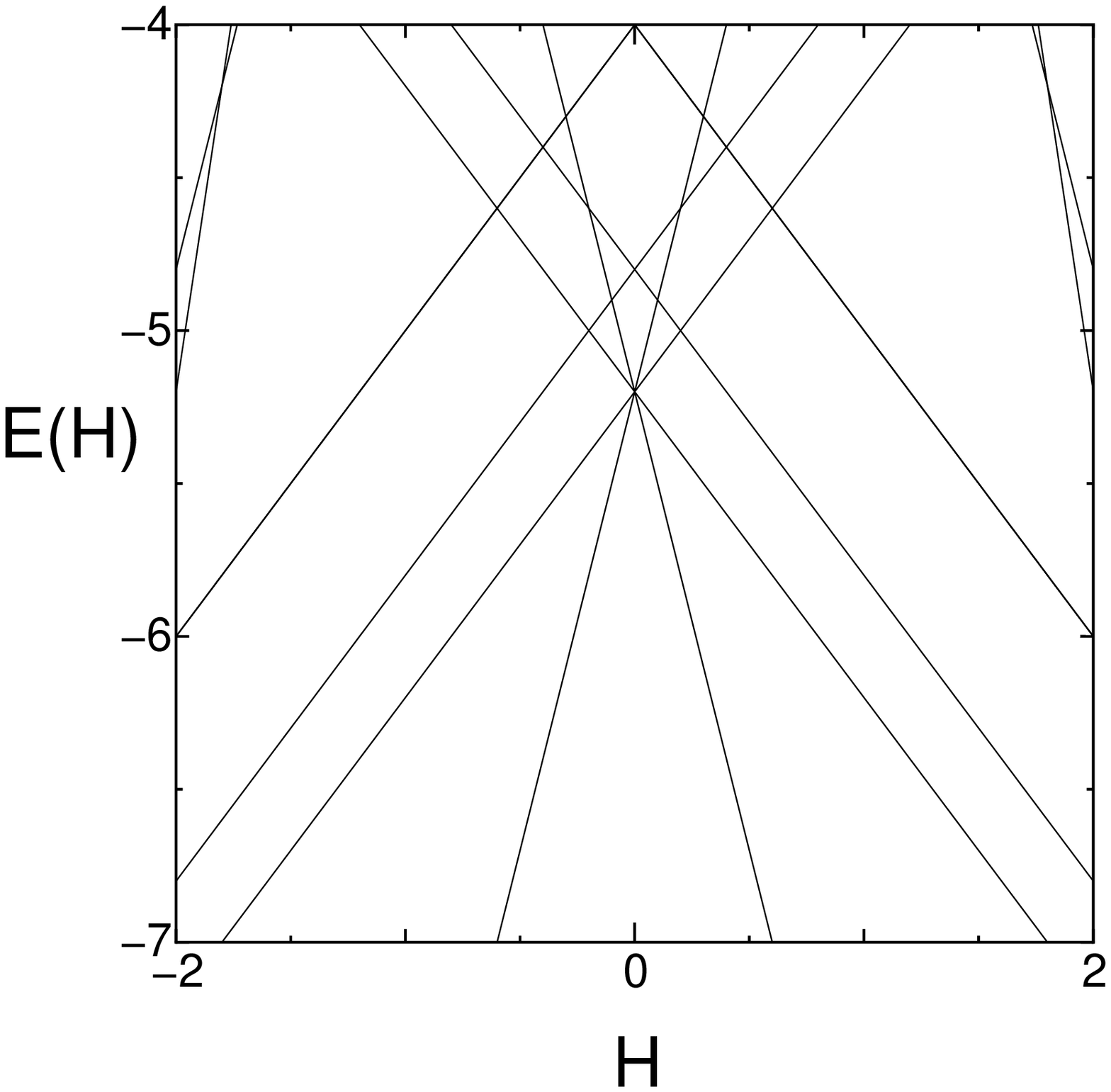} &
\epsfxsize=6cm \epsfysize=6cm \epsfbox{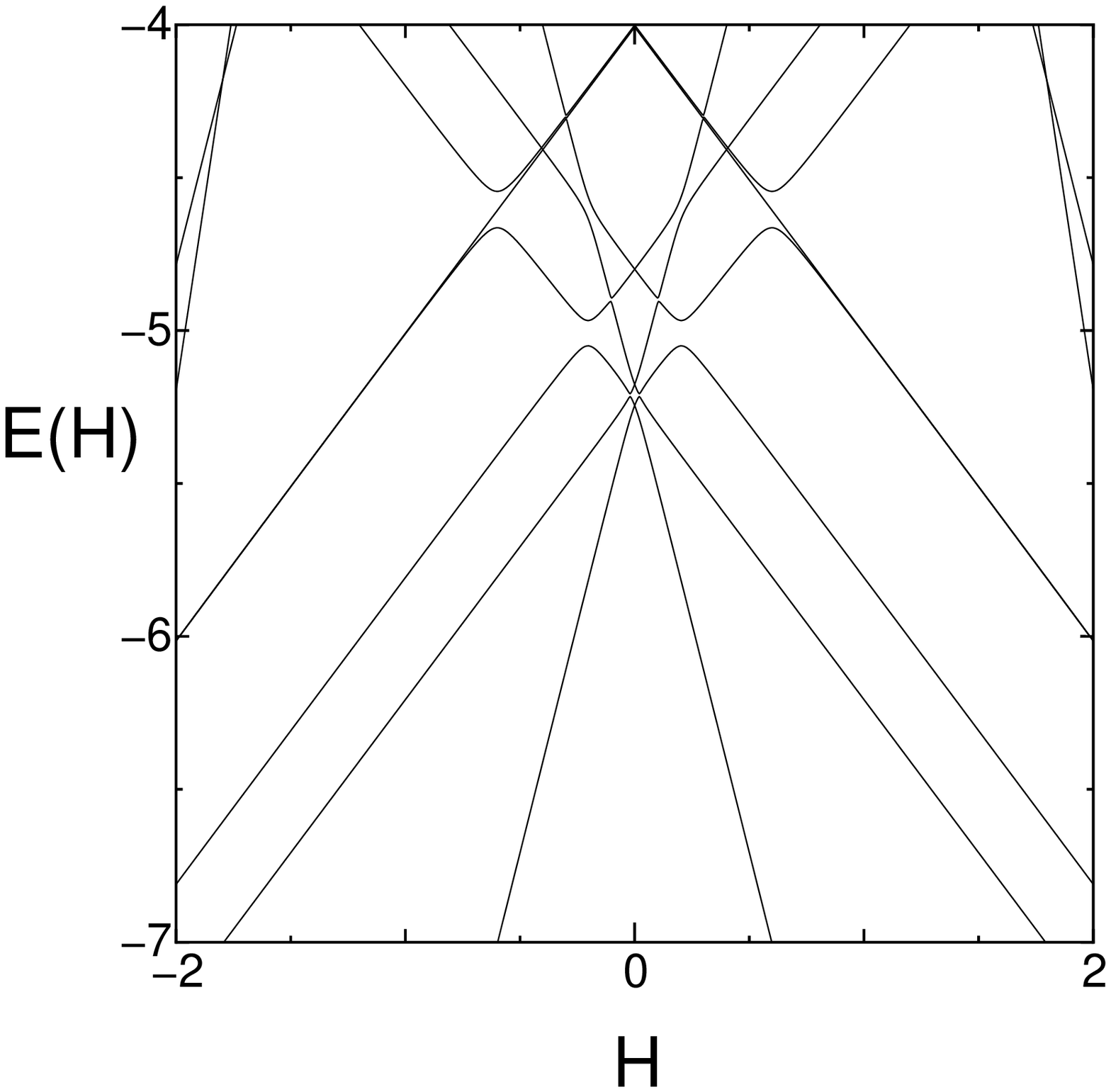} 
\end{array}
\nonumber
\end{eqnarray}
\caption{
(a) Energy structure of the pure Heisenberg model 
and (b) of the Model V with $\alpha=0.2$}
\label{spin5e}
\end{figure}

\begin{figure}
\begin{eqnarray}
\begin{array}{ll}
\epsfxsize=6.0cm \epsfysize=6.0cm \epsfbox{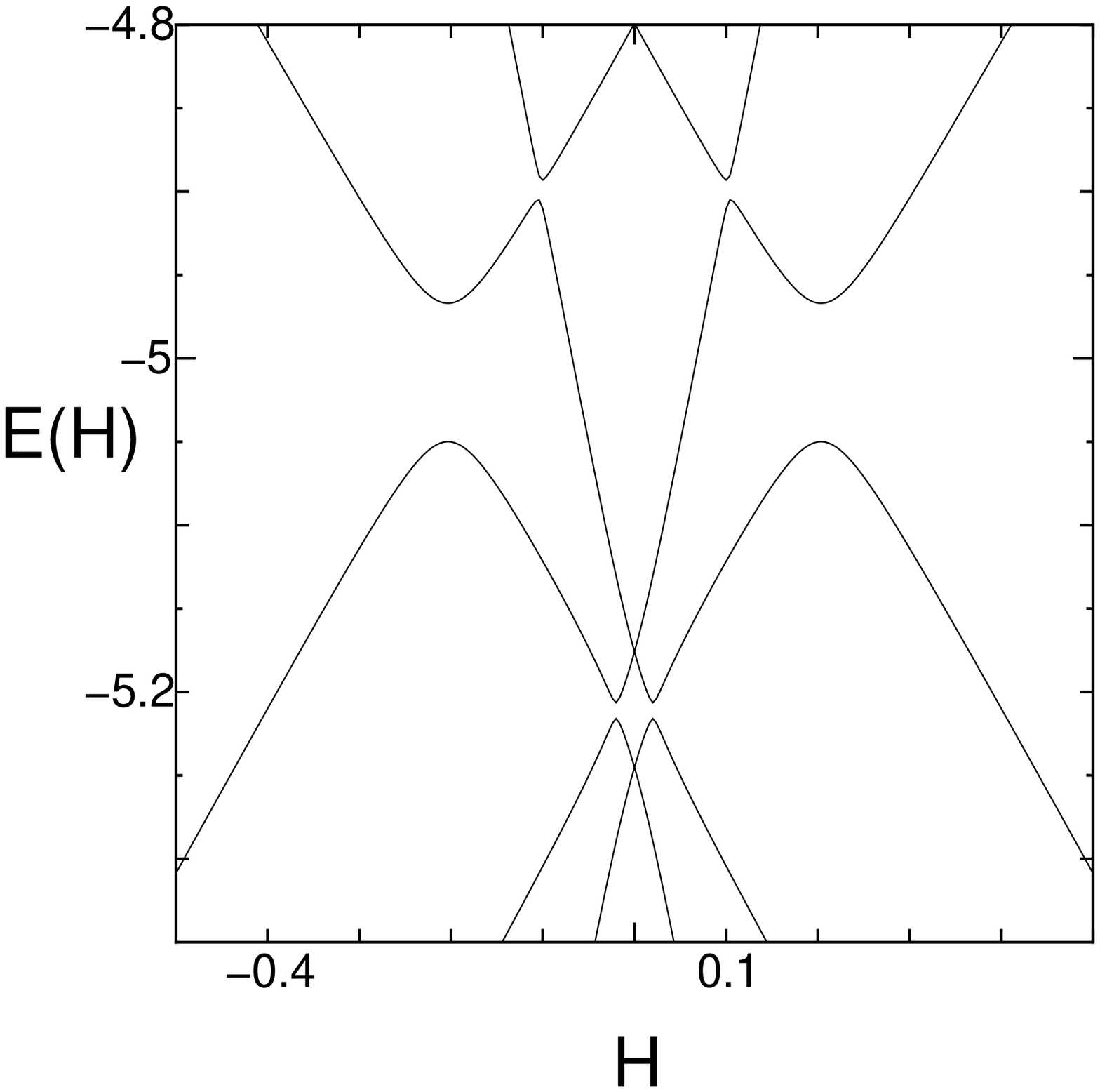} &
\epsfxsize=6.0cm \epsfysize=6.0cm \epsfbox{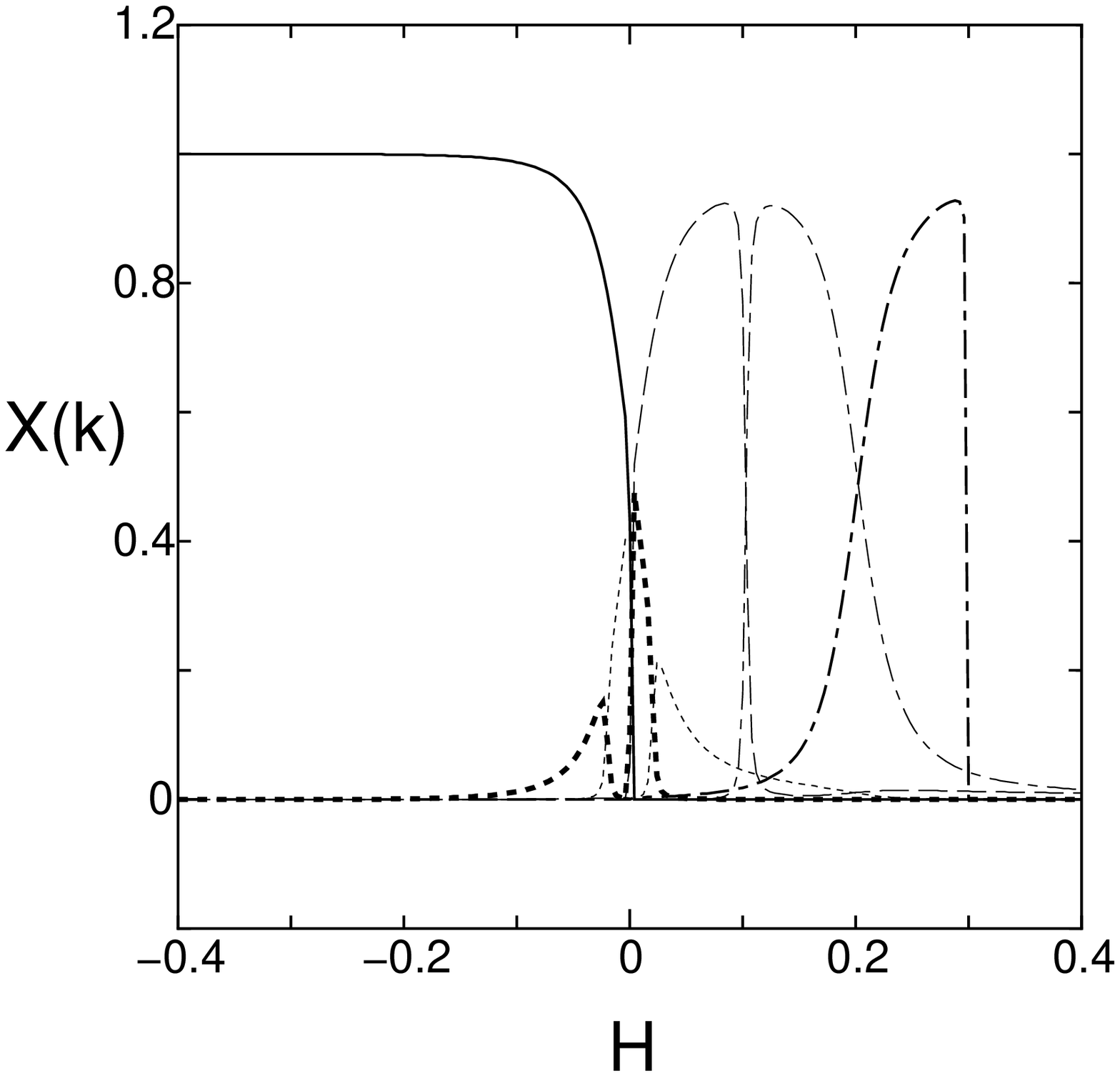}
\end{array}
\nonumber
\end{eqnarray}
\caption{(a) Zoom-up energy structure of Fig. \ref{spin5e}(b)
and (b) the overlaps of the wavefunctions}
\label{spin5x}
\end{figure}

In order to have the ground state with a higher spin  
in antiferromagnetically interacting system, 
we set $J=1,J'=0.2$. For $\alpha=0$, 
the ground state at $H=0$ is given by 4-fold degenerate $S=3/2$ states. 
The energy structures for $\alpha=0$ and $0.2$ are given in Fig. \ref{spin5e}(a) and (b),
respectively. The zoom-up around $H=0$ is shown in Fig. \ref{spin5x}(a), where
we find a complicated structure. 

The overlaps among all the four states are nonzero.
In Fig. \ref{spin5x}(b) the overlap functions are shown 
as in the same manner as eq.(\ref{OVLP}), where
the solid line, bold dotted-line, dotted-line, dashed-line, dot-dashed line, 
and dot-dot-dashed
line denote $x(k)$, $k=1,\cdots,6$, respectively.
Here we do not expect simple LZS transition.

\begin{figure}
\begin{eqnarray}
\begin{array}{ll}
\epsfxsize=6.0cm \epsfysize=6.0cm \epsfbox{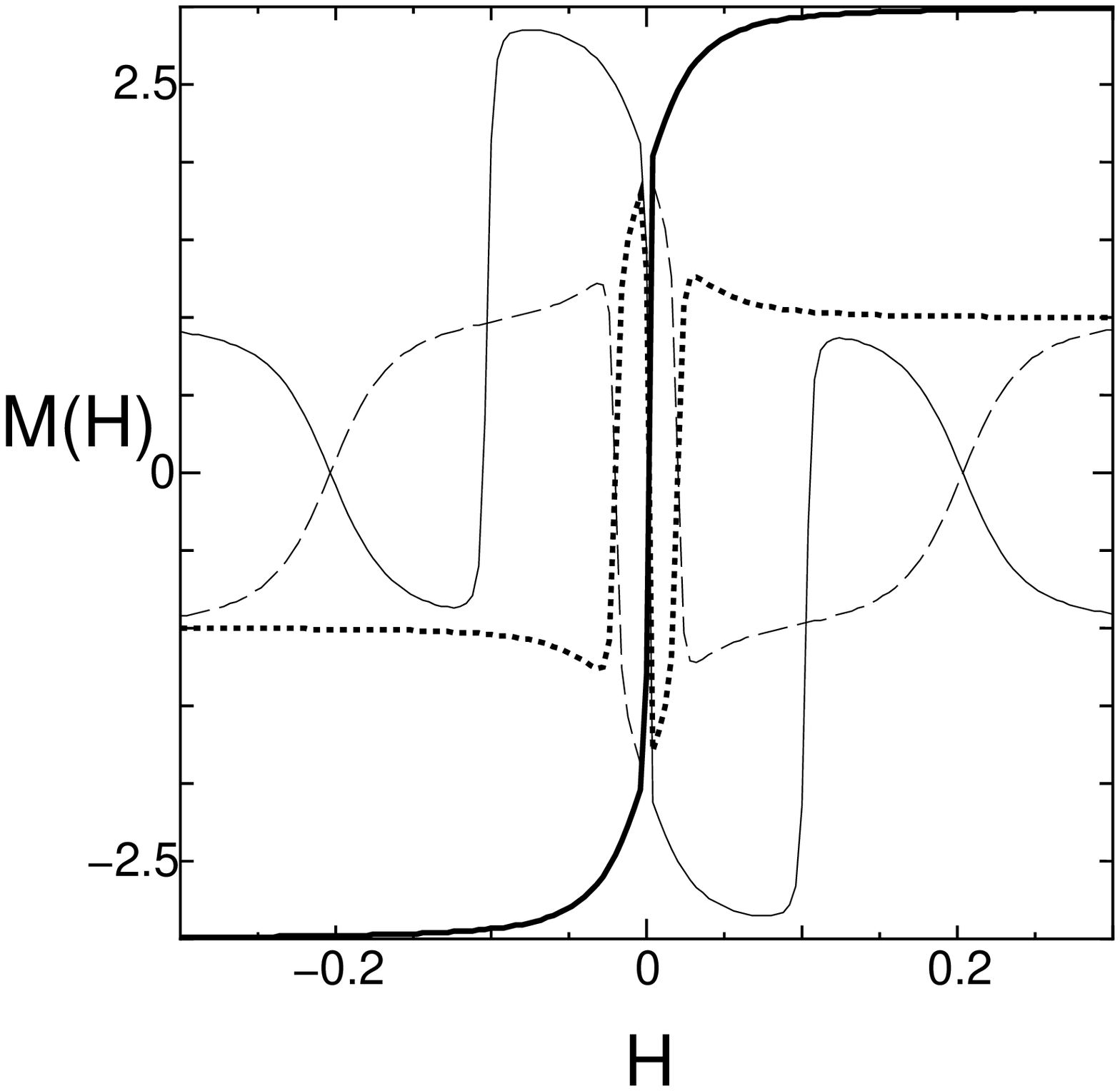} &
\epsfxsize=6.0cm \epsfysize=6.0cm \epsfbox{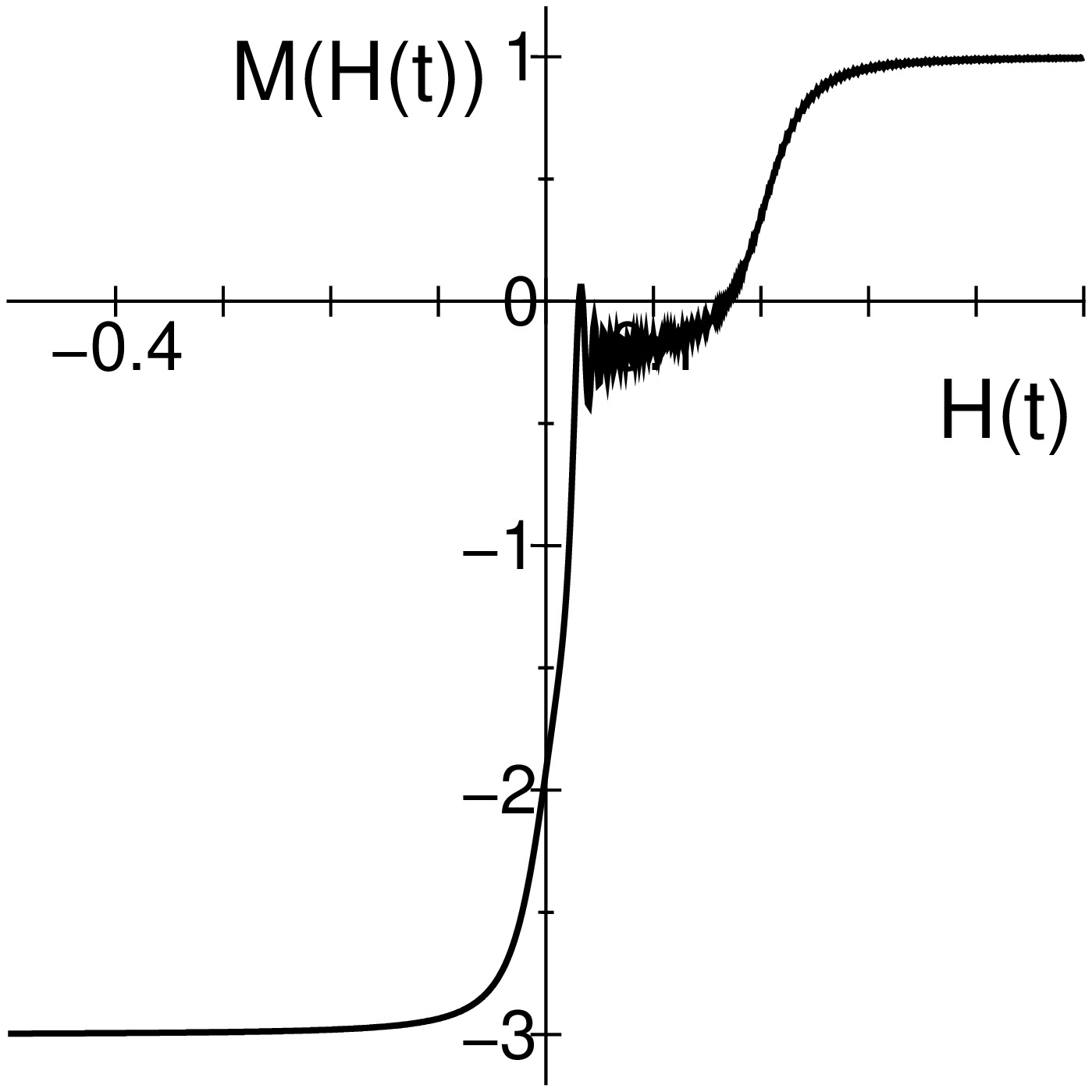}
\end{array}
\nonumber
\end{eqnarray}
\caption{(a)Magnetization of low energy levels as a function of
the field $H$, (b) Magnetization change with sweeping field $v=0.0002$}
\label{spin5m}
\end{figure}

Although we will report elsewhere detail properties of 
this non-LZS type adiabatic transition of high spin states,
we show the adiabatic magnetization processes of the present model
in Fig. \ref{spin5m}(a). If we sweep the field from a large negative value, the
adiabatic magnetization is given by the bold solid line for $H<0$.
This bold solid line jumps at $H=0$. But the ground state for $H<0$ 
adiabatically continues to the second level, and there
the adiabatic magnetization is given by the bold dotted-line. 
The second level forms an avoided level crossing with
the third level near $H=0.02$. If the nonadiabatic transition
occurs at this point, the magnetization is given by the dashed-line
instead of the bold dotted-line after this point.
Thus we expect that  
if $v<<1$ then the magnetization curve is given by
the bold dotted-line, and if $v>>1$ then by the dashed-line. 
If the field is swept with a finite speed, the
magnetization is located between the bold dotted-line and the dashed-line
depending on the speed.
In Fig. \ref{spin5m}(b), the magnetization 
change with sweeping velocity $v=0.0002$ is shown,
which is indeed between the dotted-line and the dashed-line. 

Here we adopt a peculiar example for higher spin case.
The property of the crossing would depend on the types of
perturbation.
It should be also noted here that so far we study only isotropic model
for ${\cal H}_0$. However, when we study high spin cases, the 
uniaxial anisotropy (\ref{eq1}) plays an important role, too. 
Thus, we expect a variety of types of adiabatic transition 
of the magnetization in high spin cases, which will be
reported elsewhere.

\section{Summary}

We have studied the resonant tunneling phenomena 
in half-integer spin systems with the time-reversal symmetry.
If the system has only single degree of freedom, 
the time-reversal symmetry prohibits the adiabatic change of 
the magnetization as has been pointed out.
However, if the system consists of many spins and there exist another
degree of freedom, various types of adiabatic change of the magnetization
occurs. 

In the present paper, we found degenerate (or quasi-degenerate)
avoided level crossing structures in Model III (or IV, V), where the
Landau-Zener-St\"uckelberg mechanics works. 
Appearance of avoided cross structures in half-integer spin systems
does not contradict with the Kramers theorem in all cases 
because the states belonging to different avoided level crossing structures 
degenerate at $H=0$. 

The simple LZS mechanism of the 
nonadiabatic transition does not work in some cases where more than two
levels are involved in a nonadiabatic transition (Model V). 
However an interesting
sum rule seems to hold, namely when we add the transition probabilities 
to the two neighboring states, the LZS formula gives an accurate fit to the
numerically obtained data. This issue deserves further studies as a 
possible extension of the LZS theory.

We would like to point out that the adiabatic magnetization does not
necessarily coincide with the magnetization of the ground state
even the adiabatic change of the sign of the magnetization occurs at $H=0$. 
When we sweep the field from a
large negative value where the magnetization is $-S$, we could not find any case
where the magnetization reaches to $S$ when the field becomes a large positive
value. 
Although the naive extension of (\ref{PI}) does not hold when we sweep the
field, i.e.,
\beq
\langle -S|\exp\left(-i\int_0^t({\cal H}-H(s)\sum_iS_i^z)ds\right)|S\rangle \ne 0,
\eeq
where the $H(0)=-H_0 <0$ and $H(t)=H_0$,
the present observation may suggest that
the above matrix elements is zero in the limit $H_0\rightarrow \infty$.
This problem will be studied in more general cases.

As we saw in Model I and II,
if the system has some symmetry and 
there are only two state in the same symmetry, the levels simply cross
at $H=0$ due to the time-reversal symmetry. Even in such case
the adiabatic change of the state can provide nontrivial 
behavior at $H\ne 0$ as we saw in Model II.
The thermodynamical unstable state would cause interesting 
phenomena when the system is coupled with the dissipative environments.

\section*{Acknowledgements}
We would like to thank Dr. K. Saito for valuable discussions.
The present work partially is supported by the Grant-in-Aid from 
the Ministry of Education, Culture, Sports, Science and Technology.

\appendix 
\section{Special interference on the triangle lattice}

Let us consider ${\cal H}_1$ of Model I
\beq
{\cal H}_1=\alpha(S_1^xS_2^z+S_1^zS_2^x+S_2^xS_3^z+S_2^zS_3^x+S_3^xS_1^z+S_3^zS_1^x).
\eeq 
If we consider the total spin operators
\beq
S_{\rm tot}^x=S_1^x+S_2^x+S_3^x,\quad S_{\rm tot}^y=S_1^y+S_2^y+S_3^y,
\quad S_{\rm tot}^z=S_1^z+S_2^z+S_3^z,
\eeq
${\cal H}_1$ is expressed as
\beq
{\cal H}_1={\alpha\over2}\left( S_{\rm tot}^xS_{\rm tot}^z+ S_{\rm tot}^zS_{\rm tot}^x \right)
=\alpha\left( S_{\rm tot}^xS_{\rm tot}^z+{i\over2} S_{\rm tot}^y\right).
\label{A3}
\eeq
If we apply ${\cal H}_1$ to the all-up state $|+++\rangle$, we have
\beq\begin{array}{ll}
{\cal H}_1|+++\rangle &=\alpha\left(S_{\rm tot}^x\times {3\over2}+
{i\over2}S_{\rm tot}^y )\right)|+++\rangle\\
                      &=\alpha\left({3\over 4}S_{\rm tot}^- - 
{1\over 4}S_{\rm tot}^-\right)|+++\rangle\\
&={\alpha\over2}S_{\rm tot}^-|+++\rangle.
\end{array}
\eeq
This is a symmetric state in $M=1/2$.
If we apply ${\cal H}_1$ further, we have
\beq\begin{array}{ll}
{\cal H}_1^2|+++\rangle &={\alpha^2\over2}\left(S_{\rm tot}^x\times {1\over2}+
{i\over2}S_{\rm tot}^y )\right)S_{\rm tot}^-|+++\rangle\\
                      &={\alpha^2\over2}\left({1\over 2}S_{\rm tot}^x + 
{i\over 2}S_{\rm tot}^y\right)S_{\rm tot}^-|+++\rangle\\
&={\alpha^2\over 4}S_{\rm tot}^+S_{\rm tot}^-|+++\rangle \propto |+++\rangle.
\end{array}
\eeq
Thus, the space of $M=-1/2$ can not be reached from $|+++\rangle$ by 
${\cal H}_1$. 

The present argument is directly extended to general $N$ spin system with
interactions between all spin pairs. There we find that ${\cal H}_1$ in 
(\ref{A3})
has no matrix element between the states of $M=\pm 1/2$. Thus 
the system of the type of Model I
can not adiabatically change the sign of the magnetization.

\end{document}